# Comparative Validation of Machine Learning Algorithms for Surgical Workflow and Skill Analysis with the HeiChole Benchmark

## Authors

Martin Wagner MD[a,b], Beat-Peter Müller-Stich MD[a,b], Anna Kisilenko[a,b], Duc Tran[a,b], Patrick Heger MD[a], Lars Mündermann PhD[c], David M Lubotsky BSc[a,b], Benjamin Müller[a,b], Tornike Davitashvili[a,b], Manuela Capek[a,b], Annika Reinke[d,e,f], Tong Yu MSc[g,h], Armine Vardazaryan MSc[g,h], Chinedu Innocent Nwoye MSc[g,h], Nicolas Padoy PhD[g,h], Xinyang Liu[i], Eung-Joo Lee[j], Constantin Disch[k], Hans Meine PhD[k,l], Tong Xia BSc[m], Fucang Jia PhD[m], Satoshi Kondo PhD[n,1], Wolfgang Reiter[o], Yueming Jin[p], Yonghao Long[p], Meirui Jiang[p], Qi Dou[p], Pheng Ann Heng[p], Isabell Twick[q], Kadir Kirtac[q], Enes Hosgor[q], Jon Lindström Bolmgren[q], Michael Stenzel[q], Björn von Siemens[q], Hannes G. Kenngott MD MSc[a], Felix Nickel MD MME[a], Moritz von Frankenberg MD[r], Franziska Mathis-Ullrich PhD[s], Lena Maier-Hein PhD[d,e,f,t], Stefanie Speidel PhD*[u,v], Sebastian Bodenstedt PhD*[u,v]

## Authors affiliations

[a]Department for General, Visceral and Transplantation Surgery, Heidelberg University Hospital, Im Neuenheimer Feld 420, 69120 Heidelberg, Germany

[b]National Center for Tumor Diseases (NCT) Heidelberg, Im Neuenheimer Feld 460, 69120 Heidelberg, Germany

[c]Data Assisted Solutions, Corporate Research & Technology, KARL STORZ SE & Co. KG, Dr. Karl-Storz-Str. 34, 78332 Tuttlingen

[d]Div. Computer Assisted Medical Interventions, German Cancer Research Center (DKFZ), Im Neuenheimer Feld 223, 69120 Heidelberg Germany

[e]HIP Helmholtz Imaging Platform, German Cancer Research Center (DKFZ), Im Neuenheimer Feld 223, 69120 Heidelberg Germany

[f]Faculty of Mathematics and Computer Science, Heidelberg University, Im Neuenheimer Feld 205, 69120 Heidelberg

[g]ICube, University of Strasbourg, CNRS, France. 300 bd Sébastien Brant - CS 10413 - F-67412 Illkirch Cedex, France

[h]IHU Strasbourg, France. 1 Place de l'hôpital, 67000 Strasbourg, France

[i]Sheikh Zayed Institute for Pediatric Surgical Innovation, Children's National Hospital, 111 Michigan Ave NW, Washington, DC 20010, USA

[j]University of Maryland, College Park, 2405 A V Williams Building, College Park, MD 20742

[k]Fraunhofer Institute for Digital Medicine MEVIS,Max-von-Laue-Str. 2, 28359 Bremen, Germany

[l]University of Bremen, FB3, Medical Image Computing Group, % Fraunhofer MEVIS, Am Fallturm 1, 28359 Bremen, Germany

[m]Lab for Medical Imaging and Digital Surgery, Shenzhen Institute of Advanced Technology, Chinese Academy of Sciences, Shenzhen 518055, China

[n]Konika Minolta, Inc., JP TOWER, 2-7-2 Marunouchi, Chiyoda-ku, Tokyo 100-7015, Japan

[o]Wintegral GmbH, Ehrenbreitsteiner Str. 36, 80993 München, Germany

[p]Department of Computer Science and Engineering, Ho Sin-Hang Engineering Building, The Chinese University of Hong Kong, Sha Tin, NT, Hong Kong

[q]caresyntax GmbH, Komturstr. 18A, 12099 Berlin, Germany

[r]Department of Surgery, Salem Hospital of the Evangelische Stadtmission Heidelberg, Zeppelinstrasse 11-33, 69121 Heidelberg, Germany

[s]Health Robotics and Automation Laboratory, Institute for Anthropomatics and Robotics, Karlsruhe Institute of Technology, Geb. 40.28, KIT Campus Süd, Engler-Bunte-Ring 8, 76131 Karlsruhe, Germany

[t]Medical Faculty, Heidelberg University, Im Neuenheimer Feld 672, 69120 Heidelberg

[u]Div. Translational Surgical Oncology, National Center for Tumor Diseases Dresden, Fetscherstraße 74, 01307 Dresden, Germany

[v]Cluster of Excellence "Centre for Tactile Internet with Human-in-the-Loop" (CeTI) of Technische Universität Dresden, 01062 Dresden, Germany

[1]Present address: Muroran Institute of Technology, Hokkaido, 050-8585, Japan.

*Stefanie Speidel and Sebastian Bodenstedt contributed equally to this work

## Corresponding author

Martin Wagner MD

e-mail: martin.wagner@med.uni-heidelberg.de

Postal address:
Universitätsklinikum Heidelberg
Chirurgische Klinik
Im Neuenheimer Feld 420,
69120 Heidelberg
Germany

# Abstract

PURPOSE: Surgical workflow and skill analysis are key technologies for the next generation of cognitive surgical assistance systems. These systems could increase the safety of the operation through context-sensitive warnings and semi-autonomous robotic assistance or improve training of surgeons via data-driven feedback. In surgical workflow analysis up to 91% average precision has been reported for phase recognition on an open data single-center video dataset. In this work we investigated the generalizability of phase recognition algorithms in a multi-center setting including more difficult recognition tasks such as surgical action and surgical skill.

METHODS: To achieve this goal, a dataset with 33 laparoscopic cholecystectomy videos from three surgical centers with a total operation time of 22 hours was created. Labels included framewise annotation of seven surgical phases with 250 phase transitions, 5514 occurences of four surgical actions, 6980 occurences of 21 surgical instruments from seven instrument categories and 495 skill classifications in five skill dimensions. The dataset was used in the 2019 international Endoscopic Vision challenge, sub-challenge for surgical workflow and skill analysis. Here, 12 research teams trained and submitted their machine learning algorithms for recognition of phase, action, instrument and/or skill assessment.

RESULTS: F1-scores were achieved for phase recognition between 23.9% and 67.7% (n=9 teams), for instrument presence detection between 38.5% and 63.8% (n=8 teams), but for action recognition only between 21.8% and 23.3% (n=5 teams). The average absolute error for skill assessment was 0.78 (n=1 team).

CONCLUSION: Surgical workflow and skill analysis are promising technologies to support the surgical team, but are not solved yet, as shown by our comparison of machine learning algorithms. This novel HeiChole benchmark can be used for comparable evaluation and validation of future work. In future studies, it is of utmost importance to create more open, high-quality datasets in order to allow the development of artificial intelligence and cognitive robotics in surgery.

# Keywords

# 1. Introduction

Surgical workflow and skill analysis are key technologies for the development and seemingless integration of artificial intelligence systems (AI) in the operating room (OR). These systems may increase the safety and efficiency of the operation through early context-sensitive warnings (Katić et al., 2013), OR management (Tanzi et al., 2020) and procedure time prediction (Aksamentov et al., 2017; Bodenstedt et al., 2019b), continuing surgical education (Maier-Hein et al., 2017) and professional development (Greenberg et al., 2018) by objective assessment of surgical skill and competency (Hashimoto et al., 2019; Vedula et al., 2017) as well as semi-autonomous assistance (Lalys and Jannin, 2014; Vercauteren et al., 2020). Furthermore, if today's robotic telemanipulators are to become cognitive assistance systems, which perceive their environment, interpret it according to previous experience and perform a context-aware action in a semi-autonomous way, they will have to understand the surgical workflow and learn from the skilled surgeons (Bihlmaier, 2016; Maier-Hein et al., 2017). Thus, surgical workflow and skill analysis are a prerequisite for the next generation of surgical robotics.

Given that aim, the data-driven investigation of laparoscopic surgeries is most promising. Already, the analysis of laparoscopic video data attracts increasing attention for workflow analysis (Loukas, 2018; Ward et al., 2020) as well as surgical skill assessment (Azari et al., 2019; Curtis et al., 2020). This is because laparoscopic surgery offers many advantages for the patient (Greene et al., 2007), thus being the gold standard in some routine operations, such as cholecystectomy. In consequence, this medical procedure is one of the most common operations in the United States, with approximately 300,000 operations per year (Hassler and Jones, 2020). At the same time, laparoscopy is a challenge for surgeons and assistants in their hand-eye coordination and requires a lot of training inside and outside the OR (Grantcharov et al., 2004; Schijven et al., 2005). The vascular supply of the gallbladder, for example, may be very diverse, thus confusing less experienced surgeons. In a review on the variability of the cystic artery, it has been found that in 8.9% of cases more than one artery was present and in 18.5% of cases the artery did not lie within Calot's triangle (Andall et al., 2016). In addition, there is still no conclusive solution for preventing bile duct injury (Madani et al., 2020; Mascagni et al., 2021, 2020), the most common severe complication of cholecystectomy (Michael Brunt et al., 2020). When creating AI in surgery to solve these clinical challenges, a great obstacle to train the underlying machine learning algorithms is the lack of high quality annotated data sets (Maier-Hein et al. 2021). The use of machine learning techniques has been successfully researched on the basis of annotated medical data (Topol, 2019), for example through the segmentation and recognition of tumours and metastases (Ehteshami Bejnordi et al., 2017; Golden, 2017; Heller et al., 2019), diseased atrial structures in magnetic resonance imaging (Xiong et al., 2021), pathologies in chest X-rays (Li et al., 2018; Nam et al., 2019; Singh et al., 2018; Wang et al., 2017), as well as healthy abdomen segmentation of CT and MR data (Kavur et al., 2020), or medical sensors such as for the interpretation of electrocardiograms (Strodthoff and Strodthoff, 2019; Willems et al., 1991). At the same time, research on surgical videos is comparatively underrepresented, even though the availability of surgical video data is increasing thanks to time-efficient and cost-effective recording and storage of laparoscopic surgery videos. A reason may be that the process of video annotation with meaningful information for machine learning algorithms is time-consuming and laborious. As a result, only few surgical video data sets are openly available for research, even though the publication rate for the analysis of surgical procedures has increased in recent years (Loukas, 2018). For instance, the Cholec80 data set contains videos of 80 laparoscopic cholecystectomies from a single center and annotation of surgical phase and instrument presence with a phase recognition average precision of up to 91% (Twinanda et al., 2017). Similar results have been reported on a larger multicenter data set of 1243 laparoscopic cholecystectomies, but this data is not openly available (Bar et al., 2020). The utilization of the Cholec80 data set in studies on the automatic prediction of procedure duration (Bodenstedt et al., 2019b) and autonomous annotation assistance of surgical workflow (Bodenstedt et al., 2019a; Lecuyer et al., 2020) underlines the importance and usefulness of surgical workflow recognition through machine learning. Another open data set is the Heidelberg Colorectal data set. It comprises 30 videos of three different types of laparoscopic colorectal surgeries, corresponding sensor data from medical devices in the OR and pixel-wise semantic segmentation of surgical instruments (Maier-Hein et al., 2021). The successful use of the Heidelberg Colorectal data set during the Endoscopic Vision (EndoVis) challenges 2017 (https://endovissub2017-workflow.grand-challenge.org/) and 2019 (https://robustmis2019.grand-challenge.org/) is an example for the comparative validation of machine learning algorithms to explore an optimal solution for surgical problems. Apart from laparoscopy, surgical workflow has been investigated in ophthalmology. The CATARACTS Challenge presented successful results on instrument presence detection during cataract surgery using computer vision algorithms (Al Hajj et al., 2019). Though all three data sets used clinical patient videos as the basis for their annotations, they are limited in their transferability, because they do not sufficiently reflect the diversity of clinical data in a multi-center setting limiting their representativeness. Furthermore, they focus on a limited variety of annotated features.

In contrast, the preclinical JIGSAWS data set for gesture and skill analysis contains kinematic and video data of surgical tasks with detailed action and skill annotation (Ahmidi et al., 2017). Surgical skill has a significant impact on the long-term survival of patients (Brajcich et al., 2020) and a higher skill level is associated with fewer postoperative complications and better patient outcome (Birkmeyer et al., 2013; Stulberg et al., 2020). The usage of the JIGSAWS data set in studies on the automatic assessment of technical and surgical skills (Funke et al., 2019; Ismail Fawaz et al., 2019; Wang and Majewicz Fey, 2018) emphasizes the importance of annotated data sets for the improvement of surgical training through personalized feedback (Zia and Essa, 2018). However, it does not contain real patient data.

Due to their achievements, EndoVis (https://endovis.grand-challenge.org/) (28%), Cholec80 (21%), and JIGSAWS (17%) were mentioned as the most useful publicly available datasets for surgical data science (Maier-Hein et al., 2020a), but there is still an urgent demand in the scientific community and medical device industry for high-quality data sets from laparoscopic surgery that allow a comparison of machine learning algorithms (Maier-Hein et al., 2020a). Apart from this, it is generally important to view medical recognition challenges with caution due to the lack of standardized and evidence-based quality control. For example, it is possible for later participating teams to boost their performance if the test data has been published (Maier-Hein et al., 2018). Moreover, standardized phase definitions are missing in the existing data sets, especially for laparoscopic cholecystectomy (Garrow et al., 2020; Meireles et al., 2021). Since the reproducibility of research results is an important element of science, a standardized benchmark for comparing such results is of great importance.

To counteract this deficiency and investigate the generalizability of algorithms in a multi-center setting, we present a new data set of 33 annotated laparoscopic cholecystectomies from three different surgical centers that were used in the sub-challenge "Surgical Workflow and Skill Analysis" of the international EndoVis challenge 2019 (https://endovissub-workflowandskill.grand-challenge.org/). The purpose of this study was to establish an open benchmark for surgical workflow and skill analysis by providing a state of the art comparison of machine learning algorithms on a novel and publicly accessible data set that improves representativeness with multi-centric data clinical data. Thus, our study aimed at answering the following research questions:

(1) Can the previously reported performance in recognition of surgical phase and surgical instrument be reproduced on this data set by independent researchers?
(2) What performance can be achieved for recognition tasks more difficult than phase recognition such as surgical action (often brief and subtle) and surgical skill (holistic assessment of the whole video)?

# 2. Data set

The structure of this paper follows the BIAS statement for transparent reporting of biomedical image analysis challenges (Maier-Hein et al., 2020b) and includes the structured challenge design and the reviewer checklist in the appendix.The generation of the challenge dataset is described including annotations for surgical phase, action, instrument and skill. Then, the challenge design and a description of the competing machine learning algorithms are described.

## 2.1. Data collection

The data set contains n=33 videos of laparoscopic cholecystectomies from three surgical centers in Germany with a total video length of 22 hours. The total number of cases was chosen based on annotation capacity. The operation videos at Heidelberg University Hospital (n=15) were recorded with a laparoscopic 2D camera (Karl Storz SE & Co KG, Tuttlingen Germany) with 30° optics, a resolution of 960x540 pixels and 25 frames per second. The operation videos at Salem Hospital (n=15) and the GRN-hospital Sinsheim (n=3) were recorded with the laparoscopic 2D camera ENDOCAM Logic HD (Richard Wolf GmbH, Knittlingen, Germany) with 30° optics, a resolution of 1920x1080 pixels and for the greater part 50 frames per second. Three operations at Salem Hospital were recorded with a resolution of 720x576 pixels and 25 frames per second. Every video starts at the first insertion of the laparoscopic camera into the patient's abdomen and ends with the last removal of the laparoscopic camera.

The videos were split into the training (n=24) and test (n=9) data set. In the training dataset, videos from Heidelberg University Hospital and Salem Hospital are equally represented (n=12 each). In the test data set, all three centres are equally represented (n=3 each). Assignment to training or test data set was performed with stratified random assignment. The total number of test cases was chosen to maximize the ability to generalize and evaluate while maintaining a large enough training set.

To comply with ethical standards and the general data protection regulation of the European Union, routinely collected data was used and anonymized. To this end, scenes outside the abdominal cavity, for example when the camera was pulled out for cleaning purposes, were manually censored (frames replaced were replaced with white frames) and files were renamed anonymously (HeiChole-1, HeiChole-2 etc.).

## 2.2. Data annotation

The anonymized video data were annotated with surgical knowledge by medical experts following annotation rules in the appendix. Annotation, i.e. labelling of each video frame with information about what is depicted in this frame, was performed using the video annotation research tool Anvil (Kipp, 2014). The annotation included framewise annotation of surgical phases, actions and instruments as well as skill classification for procedures and selected phases. Thus, different perspectives of the surgical activity were annotated. According to Neumuth et al. a surgical activity consists of five components (Neumuth et al., 2009), which are the functional, organizational, operational, spatial and behavioral perspectives. In this study, all perspectives exempt from the spatial were annotated. The performed action describes what is done (functional, e.g. "grasp", see paragraph "Action") and is defined as a sequence of related gestures. The performer of the action (organizational, e.g. "left hand of the surgeon") and the surgical instrument used (operational, e.g. "atraumatic grasper", see paragraph "Instrument") were annotated in relation to the exact time (behavioral, framewise annotation of the video). An example of a comprehensive annotation would be "the left hand of the surgeon performs the grasping and holding action with the atraumatic grasper at 10 minutes and 15 seconds after start of the operation".

To ensure standardization and reproducibility of annotation as well as to minimize sources of error, explicit rules were formulated for phase, action and instrument annotation. An identical procedure was followed for both training and test cases. The annotation rules are enclosed in the appendix. In order to increase the reliability of the annotation, the phases were annotated independently by three specifically instructed medical students and the surgical skill by two specifically instructed medical students. Possible error sources occurred with disagreement on the beginning or end of a phase or the skill level. Deviations were discussed and resolved by consensus between the same students.

According to the BIAS statement, a case in our data set encompassed all data for which the algorithm(s) participating in a specific challenge task produced one result. One case comprises three videos, a full laparoscopic cholecystectomy, all frames of the phase calot triangle dissection (P1) and all frames of the phase gallbladder dissection (P3), respectively. Annotations for surgical phase were one value per frame. Annotations for action were a 4D binary vector per frame indicating if the corresponding action is being performed (1) or not (0)). Annotations for instrument category were a 21D binary vector per frame, consisting of 7 instrument categories used during the EndoVis challenge, one undefined instrument shaft plus 14 unused categories reserved for future additions like further grasping instruments, with each entry indicating if the corresponding instrument category is visible (1) or not (0)). Annotations for instruments were a 31D binary vector per frame, consisting of 21 instruments used during the EndoVis challenge, one undefined instrument shaft plus 9 unused instruments reserved for future additions, with each entry indicating if the corresponding instrument category is visible (1) or not (0)). Surgical skill was annotated in five different dimensions, each ranked with integer values between 1 and 5, for each of the three videos full operation, P1 and P3.

### 2.2.1. Phase

For the surgical phases, one of seven surgical phases was assigned to each individual frame, analogous to the Cholec80 data set (Twinanda et al., 2017) following the annotation protocol in the appendix. The seven phases were preparation (P0), calot triangle dissection (P1), clipping and cutting (P2), gallbladder dissection (P3), gallbladder packaging (P4), cleaning and coagulation (P5) and gallbladder retraction (P6). The phases did not necessarily occur in a fixed order.

### 2.2.2. Action

The surgical action is the functional component of the activity a surgeon performs within a phase. Action was annotated as performed following the annotation protocol in the appendix, if any of the four actions grasp (A0), hold (A1), cut (A2) or clip (A3) occurred. Additionally, the performer of the action (organizational component) was annotated as the left hand of the surgeon, right hand of the surgeon or hand of the assistant.

### 2.2.3. Instrument

Instrument presence detection is important for surgical workflow analysis because it correlates with the current surgical phase. A total of 21 instruments (plus "undefined instrument shaft") of different types were annotated and additionally grouped into the seven categories grasper (IC0), clipper (IC1), coagulation instruments (IC2), scissors (IC3), suction-irrigation (IC4), specimen bag (IC5), and stapler (IC6). Furthermore, in different surgical centers, instruments by different vendors were used, which increases the representativeness of this data set. The stapler was not present in the test data set. For instrument presence, an instrument was annotated visible as soon as its characteristic instrument tip appeared in the image. The annotation continued when the tip disappeared later and only the shaft of the instrument remained visible. If the instrument shaft entered the field of view of the camera without its tip having been visible before, it was referred to as the "undefined instrument shaft", because even a human annotator would not be able to recognize a particular instrument due to the identically looking shafts. Three exceptions to this rule were the suction-irrigation, stapler and the clipper categories, as these instruments have characteristic shafts. Figure 1 shows sample images of the instruments from the data set.

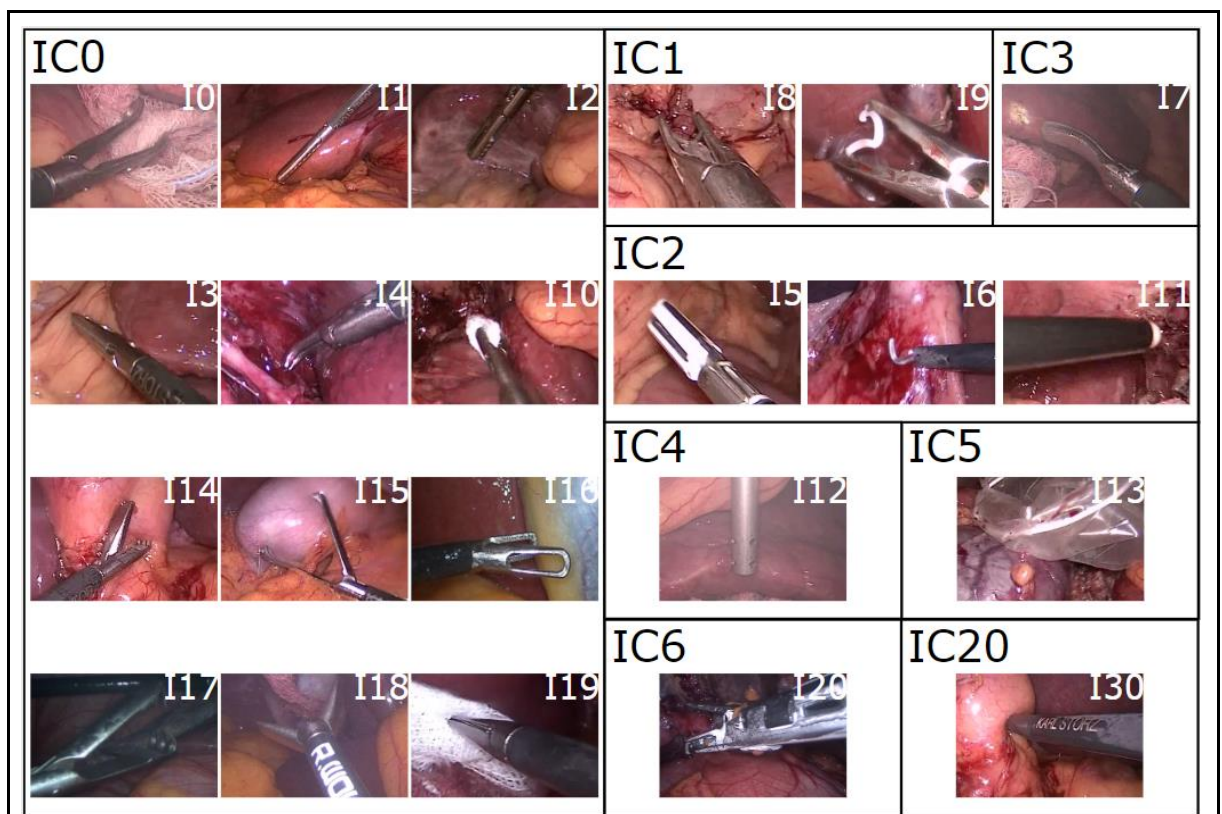

**Figure 1 - Instruments in the HeiChole benchmark.** Examples of all 21 surgical instruments plus undefined instrument shaft present in the HeiChole benchmark arranged according to the eight categories grasper (IC0), clipper (IC1), coagulation instruments (IC2), scissors (IC3), suction-irrigation (IC4), specimen bag (IC5), staper (IC6), undefined instrument shaft (IC20).
Instruments are curved atraumatic grasper (I0), toothed grasper (I1), fenestrated toothed grasper (I2), atraumatic grasper (I3), overholt (I4), LigaSure (I5), electric hook (I6), scissors (I7), clip-applier metal (I8), clip-applier Hem-O-Lok (I9), swab grasper (I10), Argon beamer (I11), suction-irrigation (I12), specimen bag (I13), tiger mouth forceps (I14), claw forceps (I15), atraumatic grasper short (I16), crocodile grasper (I17), flat grasper (I18), pointed grasper (I19), stapler (I20) and undefined instrument shaft (I30). Numbers I21 to I29 have been reserved for future additions.

2.2.4. Skill

To assess the surgical skill, the videos were scored using the modified Global Operative Assessment of Laparoscopic Skills (GOALS). It has been validated for video assessment of laparoscopic skills, including the five domains depth perception (S1), bimanual dexterity (S2), efficiency (S3), tissue handling (S4) and autonomy (Vassiliou et al., 2005). The item "autonomy" was omitted in our study, because a valid assessment based solely on intraabdominal video alone is not possible without information about what was spoken during the operation or how much assistance was provided by a senior surgeon. The difficulty of the operation (S5) was additionally annotated based on Chang's adaptation of the GOALS-score (Chang et al., 2007). Here, parameters such as inflammatory signs, adhesions and individual anatomical conditions were used to objectify the assessment of the skill. Thus, the skill assessment in this study included five ranking components. Skill was annotated for the complete operation and additionally for phases calot triangle dissection (P1) and gallbladder dissection (P3).

# 3. Methods

## 3.1 EndoVis challenge 2019

Based on our data set, 12 international teams trained and evaluated their algorithms during the EndoVis challenge 2019 (EndoVis challenge, 2021) within the sub-challenge for "Surgical Workflow and Skill Analysis" hosted in conjunction with the Medical Image Computing and Computer Assisted Intervention conference (MICCAI). The aim of this sub-challenge was to investigate the current state of the art on surgical workflow analysis and skill assessment in laparoscopic cholecystectomy on one comprehensive dataset. Specifically, the aims were (1) surgical phase recognition with high precision and recall, (2) surgical action recognition with high precision and recall, (3) surgical instrument presence detection with high precision and recall as well as (4) surgical skill assessment with a low mean absolute error. Before acceptance as a MICCAI challenge, the challenge protocol underwent a peer review process. Participants were invited to submit a Docker image and a description of the used method(s). For the submission process, participants had to register for the challenge on Synapse (https://www.synapse.org/#!Synapse:syn18824884/wiki/591922), upload their Docker container(s) to that project and then submit the appropriate version(s) to the challenge. The container(s) had to implement an interface that took a video as input, computed the appropriate challenge results and output these in CSV file(s). Only full submissions, i.e. with no results missing, for each task were considered.

The Docker images were not and will not be shared by the organizers. Each team could choose to provide their source code, though they were not required to. Only automatic methods were accepted. Participants were encouraged to provide results for recognition of surgical phase, action and instrument as well as skill assessment, but it was not required to submit in all categories and participants were free to provide results for a subset. To reduce the complexity of the challenge, not every annotation of the data set described above was used for the challenge. The action recognition did not include the recognition of the performer of an action. The instrument presence detection did not include the category undefined instrument shaft (IC20).

Participants were free to use third party public data to augment the provided training data. Submissions were "online analysis only", i.e. methods were not allowed to use information from future frames. However, in the case of skill assessment, an entire video could be used as input.

The challenge committee members were M. Wagner, S. Bodenstedt, A. Kisilenko, H. Kenngott, L. Maier-Hein, S. Speidel and B. Müller-Stich with affiliations as stated in the authors list. The award policy was to award a prize to each winner of one of the four tasks, if at least three participants entered a submission. Members of the organizing institutes were allowed to participate, but were not eligible for awards.

On May 31st 2019, the first part of the training dataset consisting of 12 videos was published, followed by the second part, also consisting of 12 videos, on August 15th 2019. With the second part, the organisers' evaluation software scripts for computing metrics and rankings were provided. The evaluation and submission period of the Docker containers was between October 1st and 7th 2019. The challenge day was on October 13th 2019. On this day, the results of all teams were presented during the EndoVis challenge meeting at the MICCAI in Shenzhen, China. Before publication of the joined paper, no results were allowed to be published.

## 3.2. Participating teams

The following sections provide a detailed description of the algorithms of the participating teams in alphabetical order. In addition, table 1 gives an overview of the different methods.

### 3.2.1. Team *CAMI-SIAT* (Phase)

The *CAMI-SIAT* team proposed a method for determining surgical phases based on Pseudo-3D residual networks (ResNet) (Qiu et al., 2017). Through the usage of a 3D convolutional network, temporal information from previous frames in an operation could be utilized directly to determine the current surgical phase. Further, they hypothesized that fusing the predictions of the Pseudo-3D residual network with prior knowledge would improve performance. For this, they determined the probability of occurrence of each phase at a given time point in the operations from the training data and applied them to the output of the network for the final prediction. No additional data was used for pre-training.

### 3.2.2. Team *CAMMA* (Phase & Action)

For the phase recognition, the *CAMMA* team utilized two different methods for image feature extraction in parallel: Inflated-3D (I3D) (Carreira and Zisserman, 2018), a 3D, and an Inception-ResNet (Szegedy et al., 2016), a 2D convolutional network. The aim of mixing 3D and 2D convolutions was to capture fine-grained temporal dynamics from the data. The convolutional networks were followed by 3 long-short-term memory (LSTM) units, one for each preceding feature extractor and one for the combined features. The predictions of the 3 LSTMs were then merged via majority-voting. During training, the binary instrument data was also used as an additional task for enhancing the results of the workflow recognition. ImageNet and Kinetics were used for pre-training.

For the action recognition, team *CAMMA* built on a ResNet (He et al., 2015) for feature extraction, which was extended with a convolutional LSTM to take temporal information into account (Nwoye et al., 2019). Following the LSTM, a combination of a convolutional layer and a fully-connected layer were utilized for higher-level reasoning on the spatio-temporal features for surgical action recognition. ResNet was pretrained on ImageNet.

### 3.2.3. Team *CareSyntax* (Instrument & Skill)

For the instrument presence detection, team *CareSyntax* utilized the approach outlined in (Vardazaryan et al., 2018), which combined a ResNet (He et al., 2015) with additional convolutional layers, so called localization maps, that helped to map features corresponding to a classification to their spatial coordinates. A spatial pooling was then utilized to determine which instrument classes were currently in use. ImageNet and Cholec80 were used for pre-training.

For the skill assessment team *CareSyntax* utilized a method based on (Funke et al., 2019). The method relied on a 3D-ResNet (Wang et al., 2016) for feature extraction. The method divided a given video into multiple segments, each segment was then fed into the 3D-ResNet. To concatenate the results of the segments, team *CareSyntax* used a fully-connected layer. The final scores were then computed by rounding the output of the network to the nearest integer. Kinetics was used for pre-training.

### 3.2.4. Team *CUHK* (Phase & Instrument)

To recognize phase and instrument, team *CUHK* implemented a multitask approach. A ResNet (Qiu et al., 2017) was used to extract visual features from a given laparoscopic image. The features were used to a) determine which laparoscopic instruments were currently visible via a fully-connected layer and b) to determine the current surgical phase via a LSTM. For the phase recognition task, the elapsed time, normalized via the average duration of all surgeries in the training data set, was concatenated to the feature representation. The output of the network was then post-processed: a median filter was applied to phase predictions with low probability and the PKI strategy (Jin et al., 2018), which takes phase order and consistency into account, was utilized to detect and correct impossible changes in phases. ResNet was pretrained on ImageNet, all networks were pretrained with Cholec80.

### 3.2.5. Team *HIKVision* (Phase & Instrument)

Team *HIKVision* utilized a ResNet (Qiu et al., 2017) for feature restriction. The ResNet was first trained for multi-task recognition of phase and instruments. The fully-connected layers for these two tasks were then replaced by a LSTM and again two output layers for phase and instrument recognition. To take phase order and consistency into consideration, PKI (Jin et al., 2018) was used to post-process the output. ResNet was pretrained on ImageNet.

### 3.2.6. Team *IGITech* (Phase & Instrument)

A ResNet (Qiu et al., 2017) was used by team *IGITech* to extract visual features from a given laparoscopic frame. A fully-connected layer was used to determine which laparoscopic instruments were located in a video frame and a support vector machine was used to recognize the current phase. It is unclear whether data was used for pre-training.

### 3.2.7. Team *Konica Minolta* (Action & Instrument)

Team *Konica Minolta* used a 101-layer ResNeXt (Xie et al., 2017) with Squeeze-and-Excitation block (Hu et al., 2019) as their base network. They attached a fully-connected layer for the instrument presence detection and one for the action recognition to the network. Furthermore, a third layer was added for detecting instrument super-classes. Here, one super-class generally consisted of a combination of instruments that regularly occur together, e.g. grasper and scissors. ResNeXt was pretrained on ImageNet.

### 3.2.8. Team *MEVIS* (Phase)

Team *MEVIS* used a ResNet (Qiu et al., 2017) for extracting visual features from the laparoscopic video frames. A LSTM was used to incorporate information from past frames. For each task, a fully-connected layer was added to the LSTM. During training, frames at the beginning of a video clip were given a weight of zero to allow the network a "warm-up phase". ResNet was pretrained on ImageNet and Cholec80. In addition, video-clips from 16 Cholec80 videos were added during training to increase the data set and video-clips were sampled stratified based on the phase-labels.

### 3.2.9. Team *NCT* (Phase, Action & Instrument)

Team *NCT* used a multi-task approach for determining from a laparoscopic video the phase, which instruments were visible and what actions were being performed. The approach used a ResNet (Qiu et al., 2017) for feature extraction, a LSTM was used to propagate information from past frames along the temporal axis and used the information for predicting current frames. For each task, i.e. phase, action and instrument presence detection) a fully-connected layer was added to the LSTM. ResNet was pretrained on ImageNet, all networks were pretrained with Cholec80.

### 3.2.10. Team *VIE-PKU* (Phase, Action & Instrument)

The approach of team *VIE-PKU* combined two approaches for feature extraction. A 2D architecture, ResNet (Qiu et al., 2017), and a 3D architecture, 3D (Carreira and Zisserman, 2018), were run in parallel to capture both spatial and temporal features. Starting from the features, three separate branches emerged, one branch built upon the features to predict the current surgical action, a second branch predicted which surgical instruments were currently visible. A third branch combined the features with the action and instrument predictions to finally determine the current surgical phase. The phase results were then post-processed using PKI (Jin et al., 2018) and also using learned intra- and inter-class probabilities. ResNet and I3D were pretrained on ImageNet, I3D was also pretrained on Kinetics.

### 3.2.11. Team *Wintegral* (Phase, Action & Instrument)

Team *Wintegral* combined several different types of models. They trained single-task models for each task, three multi-task models that combined phase recognition with instrument type detection, instrument category presence detection and action recognition respectively, as well as binary one-vs-all models that focused on a single class. For all models, a ResNet (Qiu et al., 2017) was used as a basis. The results of the different models were then aggregated and one predictor for each task was used to compute the final result. All models were pretrained on ImageNet.

| Team | Task(s) | Multi-Task | Basic architecture | (Additional) temporal component | Output component | Post-processing | Pretraining | Data augmentation | Loss function(s) | Optimizer |
|---|---|---|---|---|---|---|---|---|---|---|
| CAMI-SIAT | Phase | No | Pseudo-3D Residual Network (Qiu et al., 2017) | None (3D architecture) | Output of the network is fused with the prior probability of each surgical phase | None | None | RGB shift, brightness and contrast changes, drop-out of frames | Binary Crossentropy-loss | Adam (Kingma and Ba, 2017) |
| CAMMA | Phase | Yes (Information on instruments used during training) | Parallel I3D (Carreira and Zisserman, 2018) and Inception-ResNet (Szegedy et al., 2016) | 3 LSTMs (192, 512 and 512 units) | Majority voting to aggregate the outputs of the three LSTMs | None | ImageNet and Kinetics | None | Binary cross-entropy loss | RMSProp (Hinton et al., 2012) |
| | Action | No | ResNet-50 (He et al., 2015) | Convolutional LSTM (512 units) | A combination of convolutional layer and fully-connected layer connected to the LSTM | None | ResNet pretrained on ImageNet | None | Binary Crossentropy-loss | Momentum Optimizer |
| CareSyntax | Instrument | No | ResNet-18 (He et al., 2015) | None | Convolutional layers, spatial pooling, fully connected layer | None | ImageNet and Cholec80 | Rotation and horizontal flip | Weighted cross-entropy loss | SGD (Kiefer and Wolfowitz, 1952) |
| | Skill | No | 3D ResNet-18 | None (3D architecture) | Concatenation of results using fully-connected layer | None | Kinetics | Resizing with central cropping, random horizontal flip | Mean squared error | SGD (Kiefer and Wolfowitz, 1952) |
| CUHK | Phase | Yes | ResNet-50 (He et al., 2015) | LSTM (512 units), elapsed time | Fully-connected layers connected to | Median filter on preceding frames, Prior Knowledge | ResNet pretrained on | Cropping, flipping, mirroring and color jitter | unknown | SGD (Kiefer and Wolfowitz, 1952) |

| | | | | | as input | LSTM and instrument output Phase & Action: Fully-connected layers connected to LSTM | Inference (Jin et al., 2018) | ImageNet,all pretrained with Cholec80 | | | |
|---|---|---|---|---|---|---|---|---|---|---|---|
| | Instrument | | | | None | Fully-connected layer | | | | | |
| HIKVision | Phase | Yes | ResNet-50 (He et al., 2015) | LSTM | Fully-connected layer per task, connected to the LSTM | | Prior Knowledge Inference (Jin et al., 2018) | ResNet pretrained on ImageNet | Random crop, rotation and flip | unknown | SGD (Kiefer and Wolfowitz, 1952) |
| | Instrument | | | | | | | | | | |
| IGITech | Phase | No | ResNet-50 (He et al., 2015) | None | Support Vector Machine | | None | unknown | unknown | unknown | unknown |
| | Instrument | | | | Fully-connected layer | | | | | | |
| Konica Minolta | Action | Yes | ResNeXt-101 (Xie et al., 2017) with Squeeze-and-Excitation block (Hu et al., 2019) | None | Fully-connected layer per task | | None | ResNeXt pretrained on ImageNet | Random translation, rotation, resizing, horizontal flip and contrast changes | Binary Crossentropy-loss | SGD (Kiefer and Wolfowitz, 1952) |
| | Instrument | | | | | | | | | | |
| MEVIS | | Yes | ResNet-50 [5] | LSTM (512 units) | Fully-connected layer per task, connected to the LSTM | | None | ResNet pretrained on ImageNet and Cholec80 | Random crop and horizontal flip | Categorical Crossentropy-loss | Adam (Kingma and Ba, 2017) |
| | Phase | | | | | | | | | Binary Crossentropy-loss | |
| NCT | Phase | Yes | ResNet-50 [5] | LSTM (512 units) | Fully-connected layer per task, | | None | ResNet pretrained | None | Categorical Crossentropy-loss | Adam (Kingma and Ba, 2017) |

| Team | Task | | | | | | | | | |
|---|---|---|---|---|---|---|---|---|---|---|
| | Action | | | | connected to the LSTM | | on ImageNet, all pretrained with Cholec80 | | Dice-loss | |
| | Instrument | | | | | | | | | |
| VIE-PKU | Phase | Yes | Parallel ResNet-101 (He et al., 2015) and I3D (Carreira and Zisserman, 2018) | None (3D architecture) | Fully-connected layer connected to features | Prior Knowledge Inference (Jin et al., 2018), Inter- and intra-task correlation | ResNet and I3D pretrained on ImageNet, I3D pretrained on Kinetics | None | Binary Crossentropy-loss | unknown |
| | Action | | | | | | | | Weighted Binary Crossentropy-loss | |
| | Instrument | | | | | | | | | |
| Wintegral | Phase | Yes (Single and multi-task models) | ResNet-50 (He et al., 2015) | None | Results from different models are aggregated, one regressor per task | None | ResNet pretrained on ImageNet | Contrast changes, color jitter, center crop | Categorical Crossentropy-loss | Adam (Kingma and Ba, 2017) |
| | Action | | | | | | | | | |
| | Instrument | | | | | | | | Binary Crossentropy-loss | |

Table 1 - Overview of algorithms. The submissions of the teams participating in the EndoVis challenge are presented with components of their machine learning methods for the specific tasks.

## 3.3. Assessment Method

The properties of the dataset were analyzed with descriptive statistics mean and standard deviation (SD). To calculate results of the EndoVis challenge for the tasks of phase, action and instrument recognition, the F1-score $F1 = 2 \cdot \frac{precision \cdot recall}{precision + recall}$ over all classes was computed. The final ranking was calculated by averaging the F1-scores per video. We selected the F1-score as it takes both false positives and negatives into account. We decided to average over all classes and all videos so that each class and video would have the same weight in the final score regardless of occurence rate and length.

For skill assessment, the mean absolute error $MAE = \frac{1}{n}\sum_{i=1}^{n} |\hat{Y}_i - Y_i|$ over all criteria was evaluated and ranked. We decided to use the MAE for skill assessment, as it gives a good impression on how far predictions are on average from the reference.

To analyze the ranking stability, bootstrapping methods were applied as suggested in (Maier-Hein et al., 2018). The rankings were recalculated over 1,000 bootstrap datasets drawn from the test set to assess the variability in ranks against small perturbations. Kendall's tau (Kendall, 1938) correlation coefficient was used to quantify the changes between the original and the bootstrap rankings for all three tasks. It is ranged between -1 and 1, where a value of 1 indicates an unchanged ranking and -1 a reversed ranking. Finally, pairwise Wilcoxon signed rank tests with a 5% alpha level and adjustment according to Holm for multiple testing were computed between each algorithm pair. The tests were used to check for algorithms that were significantly superior in terms of performance compared to their competitors. The statistical analysis described above was performed using the *challengeR* toolkit, version 1.0.1, that was presented in (Wiesenfarth et al., 2021).

# 4. Results

In this section we will give a quantitative overview of the data set including the annotations, followed by the results of the EndoVis challenge 2019. Together they build the foundation for the HeiChole benchmark that will serve as a validated reference benchmark.

## 4.1. Data set for Surgical Workflow and Skill Analysis

We introduced a novel data set with 33 laparoscopic cholecystectomy videos with a total operation time of 22 hours (40.04 min ± 18.06 min per video) from three surgical centers including annotation of 250 transitions between seven surgical phases, 5514 occurences of four surgical actions, 6980 occurences of 21 surgical instruments from seven categories and 495 skill classifications in five dimensions.

### 4.1.1. Phase

There was a high variability of the individual phase lengths and occurrences (figure 2). In 28 videos (84.8%), all seven phases occurred. Each operation began with phase preparation (P0) and ended with phase gallbladder retraction (P6, 84.8 %) or cleaning and coagulation (P5, 15.2%). Phase cleaning and coagulation (P5) was omitted in five videos (15.1%). The average number of phase transitions per video was 7.6 (± 1.8 SD). The mean phase length in the data set ranged from 1.4 min ± 1.2 min (gallbladder retraction, P6) to 17.2 min ± 9.6 min (calot triangle dissection, P3).

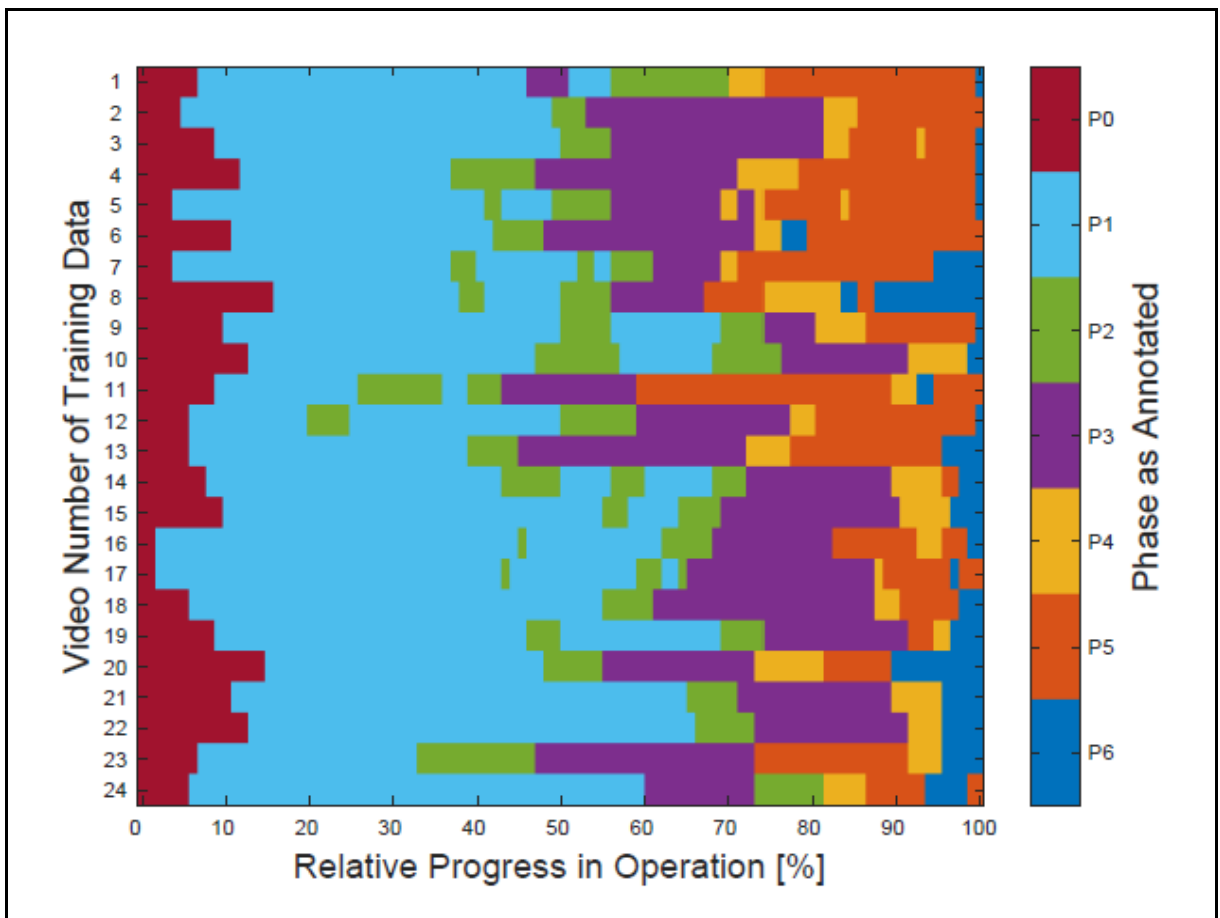

**Figure 2 - Phase distribution in the training data set.** Operation duration is normalized to 100% and phase annotations are displayed in increments of 1%. Phases displayed are: preparation (P0), calot triangle dissection (P1), clipping and cutting (P2), gallbladder dissection (P3), gallbladder packaging (P4), cleaning and coagulation (P5) and gallbladder retraction (P6).

### 4.1.2. Action

A total of 5514 actions were annotated, with an average of 167.1 ± 93.6 actions per operation. Grasp (A0) was by number of occurrences the most common action (n=2830 in total, n=28.6 ± 34.6 per operation, mean total action length per operation 9.9 s ± 11.3 s), followed by hold (A1) less often, but much longer in total (n=2124 in total, n=21.5 ± 25.9 per operation, mean total action length per operation 974.2 s ± 879.0 s). Less common as well as shorter in time were cut (A2, n=315 in total, n=9.6 ± 9.1 per operation, mean total action length per operation 7.7 s ± 10.6 s) and clip (A3, n=245 in total, n=7.4 ± 3.3 per operation, mean total action length per operation 9.6 s ± 5.4 s). Whereas the actions grasp (A0) and hold (A1) occurred throughout the complete operation, the actions clip (A3) and cut (A2) were mainly conducted during phase clipping and cutting (P2). The left hand of the surgeon, as well as the assistant hand performed only grasp (A0) and hold (A1). Cut (A2) and clip (A3) were performed by the surgeons right hand only. Mean total action length per procedure of hold (A1) was 1497.8 s ± 620.6 s (left hand of the surgeon), 68.6 s ± 56.6 s (right hand of the surgeon) and 1356.3 s ± 830.1 s (assistant hand). However, the performer (i.e. surgeon right or left hand, assistant hand) was annotated, but not published as part of the dataset.

### 4.1.3. Instrument

We found a total of 6980 instrument occurrences, with an average of 211.5 ± 134.0 instrument occurences per operation. The average instrument category presence per operation was highest for the grasper category (IC0) with 25.1 min ± 12.3 min, followed by the coagulation instruments category (IC2) with 18.1 min ± 8.4 min, specimen bag (IC5) with 4.2 min ± 3.4 min, the suction-irrigation category (IC4) with 3.8 min ± 6.7 min, the clipper category (IC1) with 1.5 min ± 0.9 min, scissors category (IC3) with 1.1 min ± 1.4 min and stapler category (IC6) with 0.03 min ± 0.16 min. Figure 8 illustrates the variation in instrument category presence depending on the progress of the operation for the test data set. For example, the grasper category (IC0) is almost continuously present in the test data set, followed by the coagulation instruments category (IC2). The categories of clipper (IC1) and scissors (IC3) are with a small proportion in the middle third of the surgeries, which corresponds with the phase clipping and cutting (P2). In contrast, the category specimen bag (IC5) has a high presence in the last third of each operation, which corresponds to the phase gallbladder packaging (P4). In the beginning of the operation, no instruments are visible, which corresponds with the preparation phase (P0), as this only includes trocar insertion and visual inspection of the abdomen. Furthermore, according to the annotation rules, the appearance of the first instrument marks the beginning of the next phase.

## 4.1.4. Skill

The skill assessment based on the ranking components ranged between the medium grade 3 and the best grade 5 (figure 3). The grades 1 and 2 have never been given. The data set contains surgeries of each level of difficulty.

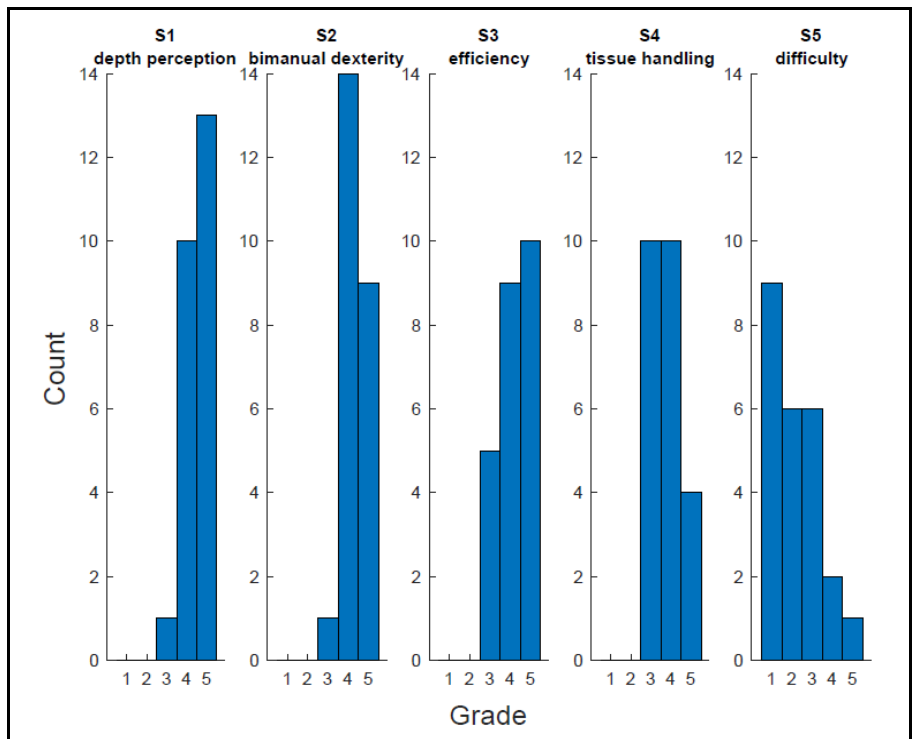

**Figure 3 - Distribution of annotated grades for surgical skill and operation difficulty in training data set.** The histograms present the annotated reference distribution of surgical skill grades within the training data set of the HeiChole benchmark. Displayed are the skill components depth perception (S1), bimanual dexterity (S2), efficiency (S3), tissue handling (S4) as well as the degree of difficulty (S5) of the operation

## 4.1.5. Differences between centers

For the reference annotation all phases and actions occurred in all the centers. The instrument presence differed per phase depending on the center, in which the operation was performed. For example, the argon beamer (I11, mean instrument presence per operation 29.83 s ± 61.05 s) was used exclusively at Heidelberg University Hospital, whereas the crocodile forceps (I17, mean instrument presence per surgery 29.72 s ± 43.85 s) was used only at Salem Hospital and GRN-hospital Sinsheim.

The cases analysed at Heidelberg University Hospital were rated to be more difficult than the cases at the smaller hospitals. There were 2 of 15 cases at Heidelberg University Hospital rated with the highest difficulty grade 5, another 2 cases with a grade 4. Only 3 cases were rated with the easiest difficulty 1. In contrast, grade 5 was not assigned to cases from either Salem Hospital or GRN-hospital Sinsheim. 8 of Salem's 15 cases were given a grade 1, as were all 3 cases from Sinsheim. Only one case from Salem was rated 4.

## 4.2. EndoVis challenge 2019

A comparison of the performance of the algorithms in the individual recognition tasks is presented in figure 4. A ranking of the algorithms together with ranking uncertainty is presented in figure 5. Both figures include results for teams CAMMA 1 and 2, which submitted after the challenge deadline and were thus not considered for the challenge awards.

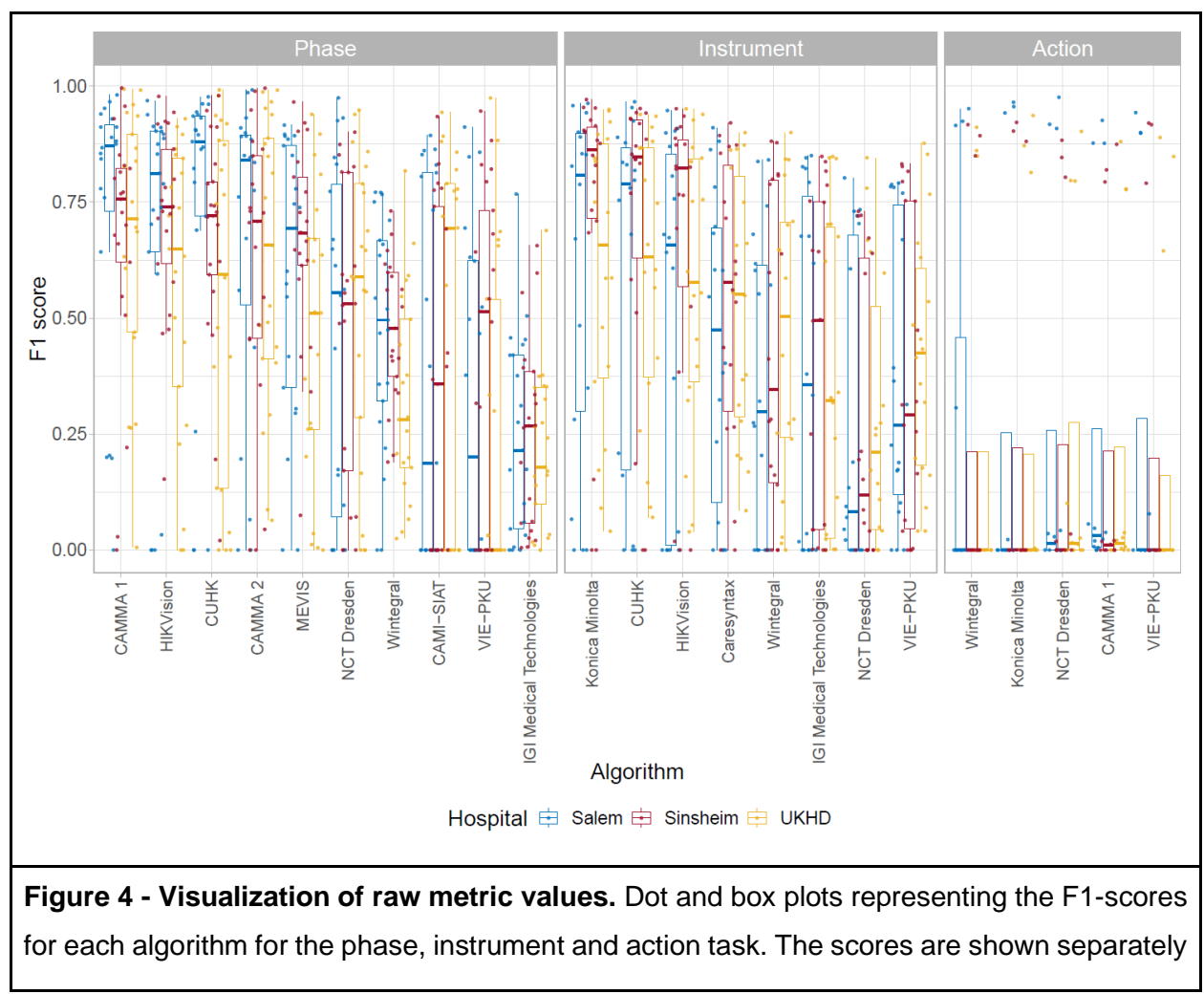

**Figure 4 - Visualization of raw metric values.** Dot and box plots representing the F1-scores for each algorithm for the phase, instrument and action task. The scores are shown separately

for each hospital from which the videos were taken. Salem is blue, Heidelberg University Hospital (UKHD) is red, GRN-hospital Sinsheim is yellow.

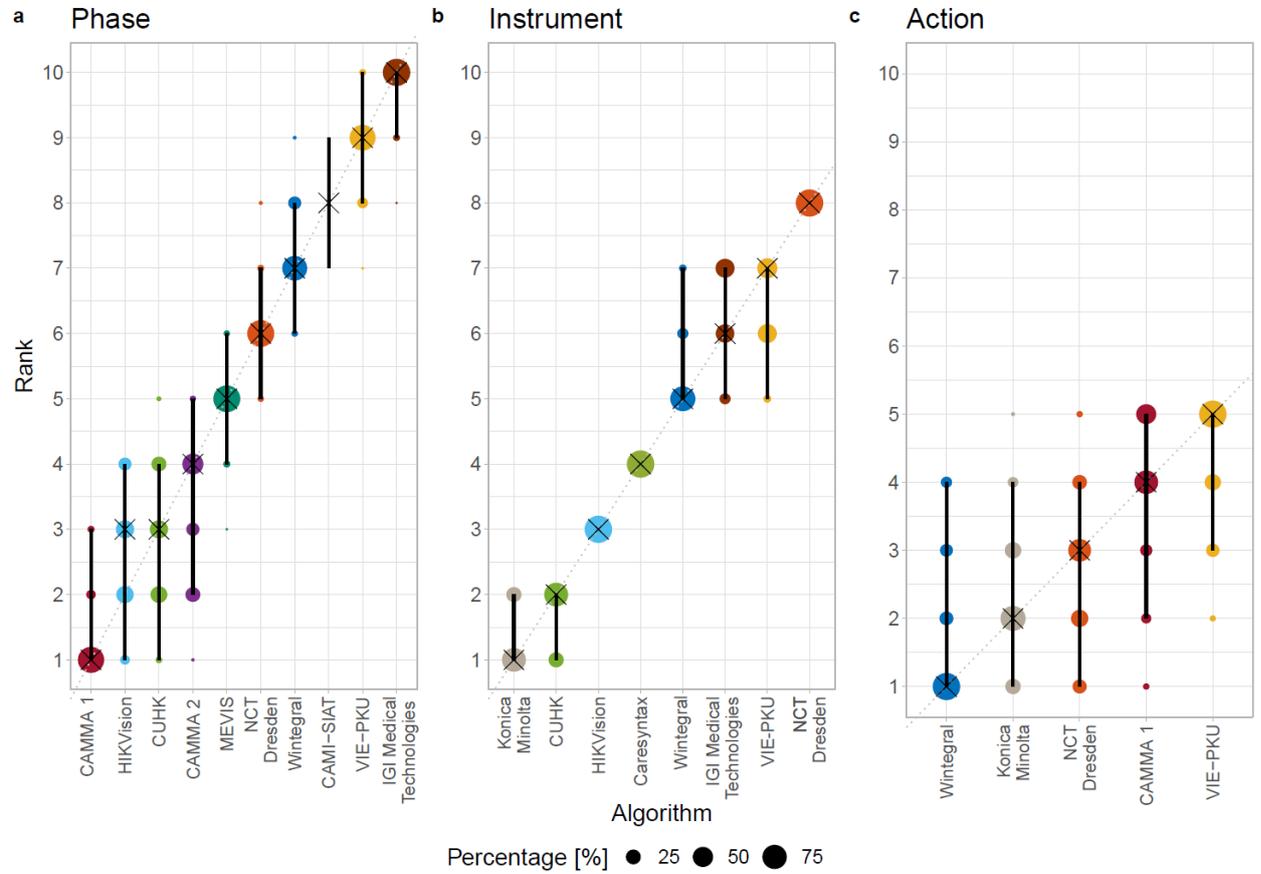

**Figure 5 - Ranking uncertainty.** Blob plots visualizing ranking uncertainty for (a) the phase, (b) the instrument and (c) the action task. The radius of the blobs represents the frequency of a rank achieved by an algorithm for 1,000 bootstrap samples. The median rank is displayed by a black cross. The 95% intervals are shown by a black line.

### 4.2.1. Phase

The results achieved in the task of phase recognition are depicted in figure 4. For phase recognition the algorithms of HIKVision and CUHK, the winning teams participating in the EndoVis challenge, achieved a F1-score up to 65.4% and 65.0%. Since the difference between the two best methods of HIKVision and CUHK was so small and the second best team achieved superior detection results in most of the videos, both were declared the winner of the challenge and received the prize. As an example, figure 6 shows a confusion matrix of the different phases for HIKVision and all participants. The highest recognition rates were achieved for phase preparation (P0, 79.4%) and calot triangle dissection (P1, 82.8%). The lowest recognition rates were achieved for phases of gallbladder packaging (P4, 50.2%) and cleaning and coagulation (P5, 54.8%).

When including submissions that were received after the challenge deadline (CAMMA 1 and 2), the best performing team CAMMA 1 and the runner-up HIKVision achieved significantly superior results to the algorithms from rank 5 to 10. No significant superiority was found for them compared to algorithms CUHK (rank 3) and CAMMA 2 (rank 4). This is also reflected by the ranking uncertainty analysis performed by using bootstrapping strategies. The frequency of ranks for each algorithm achieved across 1,000 bootstrap samples is illustrated in figure 5. It can be seen that the ranking is relatively stable, with the second and third algorithm being very close together. This is further supported by the mean Kendall's tau value for all bootstrap samples of 0.93 (median: 0.91, interquartile range (IQR): (0.91,0.96)).

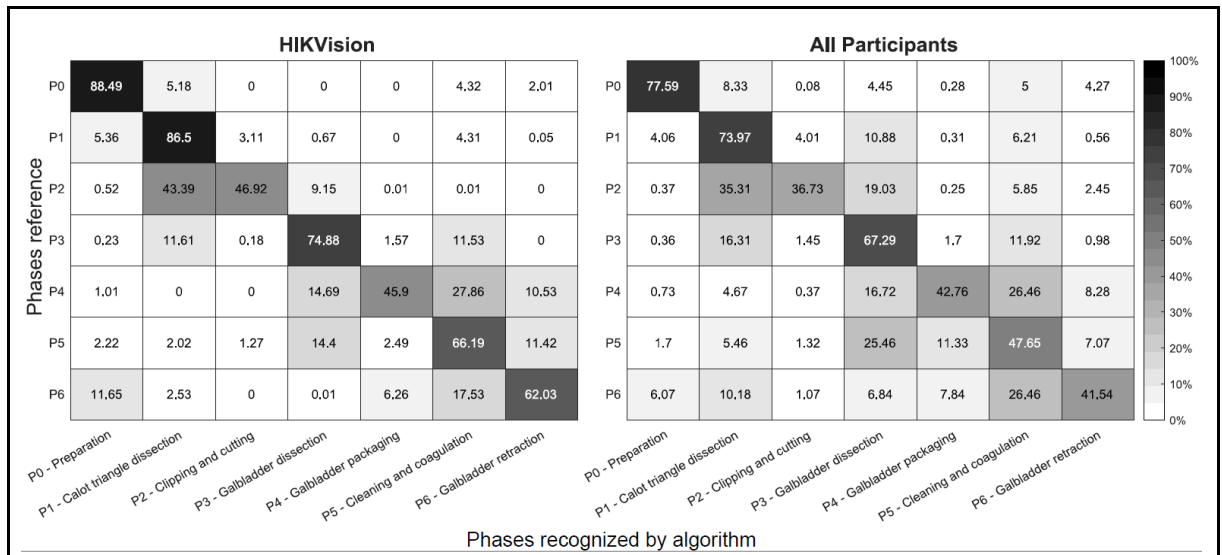

**Figure 6 - Phase recognition results.** Left: Phase recognition results for one of the challenge winners HIKVision for all test videos. Right: Results of all algorithms for all test videos. Due to statistical propagation of uncertainty, the values of the confusion matrices are only sufficiently approximated so that the cumulated lines do not add up to 100%.

### 4.2.2. Action

The task of action recognition was completed with the lowest recognition rates (figure 4). Only an F1 score of 23.3% was reached by the best team Wintegral. The action grasp (A0) occured during 1.2% (mean of all operations) of an operation, whereas the participants recognized it during 1.1% (mean recognition of all teams) of the operation. In comparison, the action hold (A1) occured during 73.3% of an operation, whereas the participants recognized it during 83.2% of the operation. The actions cut (A2) and clip (A3) occurred least frequently, each during 0.4% of an operation. Cut (A2) was recognized during 1.0% and clip (A3) during 0.3% of the operation. It is apparent that all challenge algorithms almost exclusively recognize hold (A1), whether it occurred or not. Figure 7 shows the exemplary results for the recognition of the action hold (A1) of all teams frame by frame.

None of the algorithms showed significant differences in the performance metric scores. Analyzing the ranking uncertainty revealed that the rankings were not as clearly separable as for the phase task (figure 5). The first three algorithms were closely together, therefore often exchanging their ranks. This is also reflected in a lower mean Kendall's tau of 0.66 (median: 0.60, IQR: (0.60,0.80)).

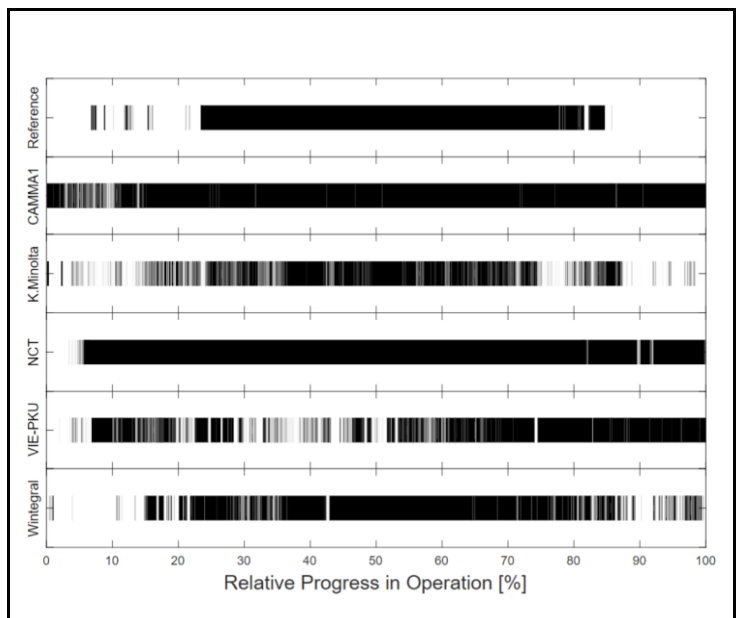

**Figure 7 – Action recognition results.** Comparison of all challenge algorithm results for recognition of action hold (A1) for test video HeiChole-25. The x-axis represents the relative time progress in the operation.

### 4.2.3. Instrument

Konica Minolta, the best team in the instrument presence detection task, achieved a F1-score of 63.8%. Their algorithm recognized 51872 instrument occurrences in the test data set, with an average of 5763.56 ± 4310.01 instrument occurences per operation. Compared to the reference, the algorithm detected a similar relative instrument presence within the operation progress, except for minor differences. For example, noticeable differences are within 10-35% of the operation progress, where the scissors (IC3) and specimen bag (IC5) were detected less frequently (figure 8). The mean results of all participants also showed considerable differences, especially in the presence detection of scissors (IC3) and specimen bag (IC5). Specimen bag (IC5) was detected continuously during the operation progress, whereas scissors (IC3) were detected less frequently than in the reference.

For the instrument task, the rankings were very stable across all bootstrap samples (figure 5). Ranks 6 and 7 were often interchanged and the first two algorithms were close together. Besides this, the ranking was quite clear with a mean Kendall's tau of 0.93 (median: 0.93, IQR: (0.93,1.00)). This was further shown by the statistical analysis. The winning team (Konica Minolta) and the runner-up (CUHK) both were superior over all other algorithms. The same trend could be seen for the other algorithms, most of them showing significant effects for their following ranks.

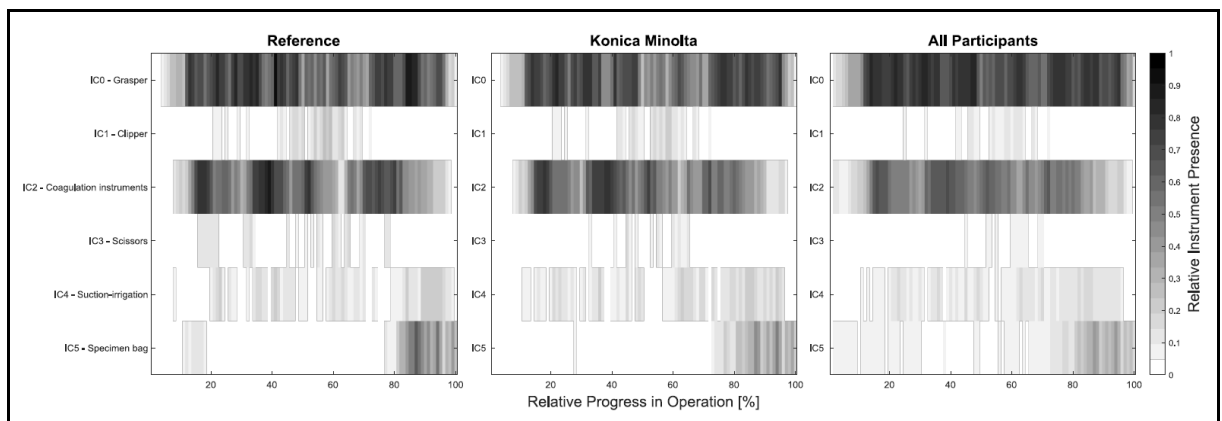

**Figure 8 - Instrument presence detection results.** For the test data set this figure compares reference annotation, averaged results of all challenge algorithms and the best challenge algorithm (Konica Minolta) for the different instrument categories IC0 to IC5 (IC6 was not present in the test data set). The horizontal axis of the graphs represent the relative time progress in the operation.

### 4.2.4. Skill

The average absolute error for skill assessment of the only participating team CareSyntax was 0.78 (n=1 team). For each surgical video, a score of 4 was given for the ranking components depth perception (S1), bimanual dexterity (S2) and efficiency, and a score of 3 for tissue handling (S4). The difficulty (S5) was rated with a score of 2 for each video.

## 4.3. HeiChole benchmark

The results of the EndoVis challenge demonstrate a comparative validation of algorithms based on our comprehensive multi task data set. New methods can be compared to these results by testing their algorithms with our data set to allow for a validated reference benchmark. The training data set will be published together with this publication. However, we decided not to publish the test data set to avoid fraud (Maier-Hein et al., 2018). To use the data set for future testing of novel algorithms against this benchmark, the following approach was implemented. To have their algorithm tested, a team must:

1. register on Synapse to register for the challenge (https://www.synapse.org/#!Synapse:syn18824884/wiki/59192). As an option, a team can be created or an existing team can be joined.
2. access the data.
3. train their algorithm on the provided training data set.
4. submit the trained algorithm.

We will then evaluate the algorithm using the test data set and report the results to the team and publish the results in an online leaderboard upon request of the submitting team. To compute a performance rank the entries will be sorted, as each task consists of a scalar metric. The current leaderboard as of the publication of this study is shown in figure 9.

| Phase recognition task | | | Action recognition task | | | Instrument presence detection task | | |
|---|---|---|---|---|---|---|---|---|
| Rank | Participant | average f1-score [%] | Rank | Participant | average f1-score [%] | Rank | Participant | average f1-score [%] |
| 1 | CAMMA 1* | 68.78 | 1 | Wintegral | 23.28 | 1 | Konica Minolta | 63.82 |
| 2 | HIKVision | 65.38 | 2 | Konica Minolta | 22.83 | 2 | CUHK | 62.95 |
| 3 | CUHK | 64.98 | 3 | NCT Dresden | 22.62 | 3 | HIKVision | 58.20 |
| 4 | CAMMA 2* | 63.60 | 4 | CAMMA 1* | 22.10 | 4 | Caresyntax | 50.13 |
| 5 | MEVIS | 57.30 | 5 | VIE-PKU | 21.75 | 5 | Wintegral | 41.59 |
| 6 | NCT Dresden | 49.00 | | | | 6 | IGI Medical Technologies | 38.86 |
| 7 | Wintegral | 42.47 | | | | 7 | VIE-PKU | 38.47 |
| 8 | CAMI-SIAT | 38.65 | | | | 8 | NCT Dresden | 27.45 |
| 9 | VIE-PKU | 33.29 | | | | | | |
| 10 | IGI Medical Technologies | 23.93 | | | | | | |

**Figure 9 – Leaderboard: Rankings for the phase, instrument and action task.** The table shows the ranking as of the publication of this study for the submissions of all participants according to the task. Participants marked with * have entered after the submission deadline of the EndoVis challenge.

# 5. Discussion

## 5.1. HeiChole Benchmark

Our study introduces the HeiChole benchmark by presenting a novel open data set for surgical workflow and skill analysis together with the results of the EndoVis challenge 2019, sub-challenge for surgical workflow and skill analysis, and its comparative validation of machine learning algorithms for automatic phase, instrument, action and skill recognition in laparoscopic cholecystectomy. In the following sections we will discuss achievements and limitations of the data set and the performance of the algorithms for each task. Finally, we will outline potential clinical applications as well as future research directions for surgical workflow and skill analysis.

## 5.2. Data set quality

One achievement of the data set is a reliable reference annotation reflecting the variability of surgical video data in different hospitals for the commonly performed operation laparoscopic cholecystectomy. The varying resolutions and frame rates of the recorded videos illustrate the high variability of the recording of even a standard operation within different centers. Moreover, as described earlier, even the instruments used in the data set vary between centers and to a certain extent even within one of the centers. This underlines the necessity to train recognition algorithms on data sets of various hospitals in order to increase the performance and thus the applicability and generalizability. The phase annotation was based on the established Cholec80 data set (Twinanda et al., 2017). This addresses the problem of high heterogeneity of phase definitions reported by different groups (Garrow et al., 2020). Furthermore, we present the first data set with surgical action annotated frame-by-frame in a clinical setup. Previously, for example in the JIGSAWS data set, gestures were annotated only in an experimental setup (Ahmidi et al., 2017). In addition, the data set is supplemented by a skill assessment based on the established GOALS score (Vassiliou et al., 2005). The reliability of phase and skill annotation is enhanced by multiple raters who annotated according to precise rules in an annotation protocol and resolved disagreement by consensus. Finally, the data set is published to make it transparent, open and reusable for the scientific community.

Despite these accomplishments, there are also some limitations to the data set. First of all, a limitation of reference annotation is the difficulty in sufficiently addressing interindividual differences in patient anatomy by annotation rules. Within the HeiChole benchmark the phase clipping and cutting (P2) was annotated for only one specific artery and one cystic duct, but not for additional vessels such as small veins. Consequently, the HeiChole benchmark does not reflect the potential diversity of the vascular anatomy as previously described (Andall et al., 2016) in order to limit the complexity of the recognition tasks further. Furthermore, consistent action and instrument annotation was difficult when the performing instrument was pivoted out of the camera's view for example by repeatedly disappearing from camera view during the action hold (A1) or the instrument was no longer visible due to smoke generated by coagulation. To address this, the reliability of annotations, for example of instrument presence or electric coagulation, could be improved by the additional acquisition of medical device sensor data (Maier-Hein et al., 2021).

The effects of different resolutions and framerates on the performance of machine learning algorithms are difficult to foresee. While a higher framerate could probably be beneficial, increased resolution could introduce more noise to the processed image, thereby decreasing the algorithms' performance. Furthermore resolution differences can also act as confounders. In contrast to this, it is also possible that the image data added at a higher resolution is beneficial. Further research is needed to assess these effects accurately. Consequently, algorithm performance should not be directly compared using data sets with different parameters such as resolution and frame rate without further reflection on the effects of the parameters (Roß et al., 2021).

## 5.3. Surgical phase recognition

In comparison to the previously achieved average precision of 91% for automatic phase recognition on the Cholec80 data set (Twinanda et al., 2017), the best participating team in this EndoVis challenge reached a F1-score of 65%. One possible reason for the lower results is that, in contrast to the Cholec80 data set, there was no rigid phase order for the HeiChole data set, which makes the recognition of the phases more complex. Furthermore, phases such as clipping and cutting (P2) and gallbladder packaging (P4) proved to be more challenging to recognize than others, because instruments typical for these phases, such as scissors for P2 or the specimen bag for P4, were also briefly used in previous or subsequent phases. In surgical reality, there are many variations in the workflow, especially for the more difficult cases that at the same time could benefit more from the help of computer assistance. Thus, the reflection of variations is important for the future clinical translation of the results. Another important factor in improving the results is increasing the size of the open data set to enhance the training of machine learning algorithms. For example, phase recognition by machine learning on a data set of 1243 laparoscopic cholecystectomies was published, but only hyperparameters, but not the training data were made publicly available (Bar et al., 2020).

From a technical aspect, it is interesting to note that the top three teams (CAMMA, CUHK and HIKVision) all used a combination of CNNs with a recurrent network, here in all cases a LSTM. While it can be expected that these methods outperform approaches that do not take temporal information into consideration (e.g. Wintegral), it is of interest to note that the recurrent approaches also generally outperform methods based on 3D convolutions. This could be explained by the fact that, at least in theory, the approaches utilizing a recurrent architecture can recall information from any previously seen frame, while the hindsight of the 3D convolution-based networks is limited by the size of the convolutional kernels used and the amount of layers to more of a local scope. The top-performing team, CAMMA, actually utilizes and merges these global and local scopes, showing that both have their merit in phase recognition.

It can be observed that the top methods performed best on the data collected from Salem and worst on the data collected from UKHD. The results on the data from Sinsheim, which was not included in the training data, are generally between those for the other two centers. This can be explained by the fact that the cases from UKHD tended to be more complicated in nature and thus have a higher variance in surgical progression. This generally makes recognizing the correct phase more complicated, while the other two centers tended to treat more standard cases. This observation leads us to conclude that, while it is important to include data from different centers in datasets, because tools and surgical guidelines tend to vary, it is at least equally important that the data should reflect the variance in patient anatomy and case difficulty.

## 5.4. Surgical action recognition

The algorithm of the leading team Wintegral has almost exclusively detected the action hold (A1), which is the most common action in the data set and led to a class imbalance. Detailed analysis of the performance within the HeiChole-25 video demonstrates that this action was assigned to frames that were annotated as such in the reference, but also to frames that were assigned to other actions or where no action at all had taken place. This observation is consistent with the percentual average of the recognized actions over the entire test data set. Thus, the comparably high performance of phase recognition could not be reproduced for the more difficult recognition task of surgical actions. This may have different reasons. In this study, surgical actions during a cholecystectomy were annotated and recognized for the first time in the literature. Neither we nor the challenge teams could utilize reference results or training data sets of other research groups and thus broke new ground. Additionally, analogous to the instrument presence, several, mostly brief and subtle, actions may occur during the same frame. This complicates the recognition process. The results may surely be optimized by further research and larger data sets for algorithm training.

While the participating teams all used varying methods for the action detection, ranging from vanila feed-forward neural networks to 3D convolutional network and to recurrent networks, the resulting performances were quite similar. The submitted methods were all able to similarly well detect the action "Hold" (F1 scores ranging from 0.84 to 0.91), but could not reliably detect any other action. The average F1 scores for each action apart from "Hold" for each method was well below 0.03. Since "Hold" was the most common action in the dataset, we conclude that the current training dataset did not adequately mirror the variance in the other actions.

## 5.5. Instrument presence detection

Overall, the task of instrument presence detection can be as challenging as phase recognition.In addition, several instrument categories can be used within a single frame. Also the detection of the small instrument tips is impeded by the non-static camera, quickly changing perspectives and the resulting motion blur, whereas phase recognition can utilize information from the whole surgical scene including the patient anatomy.

Furthermore, the analysis of the instrument presence distribution in the phases within the reference test data set highlights that certain instrument categories were used more frequently during certain phases according to their functionality leading to a class imbalance. This conclusion can serve as a reference point for further research on automatic workflow recognition, but should be supported by other sources of information, such as real-time sensor data, as variations may occur. The instrument category grasper (IC0) is the only instrument used in phases P1 through P6 with a high proportional presence. This is little surprising, since the instruments of this category are universally applicable for trivial grasping as well as for blunt dissection. The other instruments also fit, according to their function, into the phases in which their highest proportional presence was detected. For example, the category scissors (IC3) are mainly used during clipping and cutting (P2), when they are required to cut the vessels, and to a lesser extent during calot triangle dissection (P1), for the dissection of highly adherent tissues. Similarly, the instruments from the category clipper (IC1) were mainly used during clipping and cutting (P2), with small variations to the previous and subsequent phases, for example, for clipping varying vascular supply to the gallbladder.

The results that the teams submitted show that for instrument presence detection, the temporal component does not seem to be as important as for tasks such as phase recognition. The highest performing method (Konica Minolta) for example did not include any temporal information at all, instead opting to explicitly learn the co-occurrences of different types of surgical instruments. Data augmentation also seems to help increase accuracy on the given task, as it can be noted that the top 5 methods all used some form of data augmentation, while the bottom 3 methods did not.

It is interesting to note that most submitted methods actually performed best on data from Sinsheim, which was not represented in the training data, and very closely followed by Salem. The make of the tools used in Sinsheim are identical to the ones in Salem, meaning that models that perform well on Salem data should generalize to Sinsheim. In the videos from UKHD, a much larger variety of instruments are used, increasing the difficulty of determining the right class.

## 5.6. Skill assessment

The results of the challenge algorithm for skill assessment do not sufficiently reflect the reference, because every video was repeatedly rated with the same score for each ranking component. However, these results are not representative as only one team participated in this task. Here, the difficulty in skill assessment is to provide a holistic evaluation considering many aspects of the video. For this reason, the reference annotation was also performed by two independent raters in order to achieve more objective results. The raters did not use the full range of the scores in the final grading. The lack of exploitation of the lower grade scores was already noted in earlier works on surgical skill rating (Doyle et al., 2007). It can be attributed either to the best possible effort of the surgeons and thus the selection of the data, or an error of central tendency, in which the raters distribute average ratings regardless of performance.

## 5.7. Reducing bias in the challenge

The introduction and validation of the HeiChole benchmark succeeded by means of the challenge design of the EndoVis challenge. This allowed the comparison of the results of several different algorithms submitted by international teams, demonstrating the usability of the HeiChole benchmark as an validation tool for phase, instrument, action and skill recognition. As biomedical challenges become increasingly important for the evaluation and validation of methods for surgical data science, quality control is of utmost relevance to ensure reproducibility and comparability of the results (Maier-Hein et al., 2018). For example, this was addressed in the EndoVis challenge by providing detailed information about the data collection and processing in response to the challenge design questionnaire required by MICCAI 2019 for submission. In order to avoid possible fraud by the challenge participants, the test data set and the corresponding reference annotation was not published. This prevents any manipulation of the algorithm performance, so that all (future) teams participate under equal conditions.

Furthermore, there are some technical parameters to consider, which affect the challenge performance of the teams. Thus, the effects of different resolutions and framerates on the performance of machine learning algorithms are difficult to foresee. While a higher framerate could probably be beneficial, increased resolution could introduce more noise to the processed image, thereby decreasing the algorithms' performance. In contrast to this, it is also possible that the image data added at a higher resolution is beneficial. Further research is needed to assess these effects accurately. Consequently, challenge algorithm performance should not be directly compared using data sets with different parameters such as resolution and frame rate without further reflection on the effects of the parameters.

Finally, it is important to acknowledge that the plots shown in this paper should only be considered as an approximate representation of the data shown. Comparing videos with different numbers of frames makes it necessary to evaluate the progress of the operations in relative values, so rounding is necessary to achieve an integer number of frames in a given percentage range. Therefore, at the end of each operation, a number of frames is being omitted in the visualisation. This number of frames is always less than one percent. These problems are inevitable when trying to visualise large and inhomogeneous data sets, however, they should be considered when interpreting the challenge results.

## 5.8. Potential clinical applications

The aim of surgical data science is to support surgical procedures in a way that they can be performed with consistently high-quality results despite an increasing lack of physicians and regardless of demanding shifts and night work. Thus, key clinical applications may be decision and risk management support, context-aware assistance and the improvement of surgical training and education by evaluating surgical skills and feedback (Maier-Hein et al., 2017). These and similar algorithms may facilitate general operating room coordination by predicting the remaining time of the operation (Bodenstedt et al., 2019b). Algorithms may also help to locate critical events in surgical videos and document safety measures for future reference or continued surgical education (Korndorffer et al., 2020; Mascagni et al., 2021). Through successful real-time workflow analysis, autonomous robotic systems can learn from the surgeon and their own experience to adapt to individual conditions and successfully support during surgical tasks (Wagner et al., 2021). These surgical robots have the potential to improve not only the functional outcome and efficacy, but also the accessibility of surgical care in remote areas as envisioned in previous works on surgical robotic automation (Shademan et al., 2016).

## 5.9. Future Research Directions

In future studies, the applicability of automatic phase and action recognition, instrument presence detection and skill assessment should be investigated for more complex surgical procedures, such as esophageal or pancreatic surgery for cancer. Larger data sets are essential to improve the performance of the algorithms, ideally with addition of medical device sensor data to complement manual reference annotation. Also, to speed up annotation processes the development of time and cost effective annotation tools should be realized. The machine learning algorithms should incorporate surgical knowledge, such as correlation of certain instruments and phases.

# 6. Conclusion

Surgical workflow and skill analysis are promising technologies to support the surgical team, but are not solved yet, as shown by our state of the art comparison of machine learning algorithms on the novel HeiChole benchmark. The continued creation of open, high-quality data sets is of utmost importance in order to allow the development of accurate and robust machine learning algorithms as a foundation for artificial intelligence and cognitive robots in surgery.

# User Notes

The data set was published under a Creative Commons Attribution-NonCommercial-ShareAlike (CC BY-NC-SA) license on Synapse (https://www.synapse.org/#!Synapse:syn18824884/files/), which means that it will be publicly available for non-commercial usage. Should you wish to use or refer to this data set, you must cite this paper. The licensing of new creations must use the exact same terms as in the current version of the data set.

# Conflicts of interest

M.Wagner., B.P.Müller-Stich, S. Speidel and S. Bodenstedt worked with medical device manufacturer KARL STORZ SE & Co. KG in the projects "InnOPlan", funded by the German Federal Ministry of Economic Affairs and Energy (grant number BMWI 01MD15002E) and "Surgomics, funded by the German Federal Ministry of Health (grant number BMG 2520DAT82D and BMG 2520DAT82A). Lars Mündermann is an employee of KARL STORZ SE & Co. KG. A. Reinke works with the Helmholtz Imaging Platform (HIP), a platform of the Helmholtz Incubator on Information and Data Science. S. Kondo was an employee of Konica Minolta Inc. when this work was done. Wolfgang Reiter is an employee of Wintegral GmbH, a subsidiary of medical device manufacturer Richard Wolf GmbH. I. Twick, K. Kirtac, E. Hosgor, J. Lindström Bolmgren, M. Stenzel and B. von Siemens are employees of Caresyntax GmbH. Felix Nickel received travel support for conference participation as well as equipment provided for laparoscopic surgery courses by KARL STORZ SE & Co. KG, Johnson & Johnson, Intuitive Surgical, Cambridge Medical Robotics, and Medtronic.

## CRediT authorship contribution statement

**M. Wagner, A. Kisilenko:** Conceptualization, Methodology, Validation, Writing - Original Draft, Writing - Review & Editing, Formal analysis, Data Curation, Investigation, Challenge organisation, Resources, Project administration

**S. Bodenstedt:** Conceptualization, Methodology, Validation, Writing - Original Draft, Writing - Review & Editing, Formal analysis, Data Curation, Challenge organisation, Visualization, Data Analysis, Software, Resources, Project administration

**D. Tran, P. Heger, L. Mündermann, F. Nickel, M. von Frankenberg, F. Mathis-Ullrich:** Resources, Writing - Review & Editing, Data Curation

**D.M. Lubotsky, A. Reinke:** Visualization, Formal analysis, Resources, Writing - Review & Editing

**B. Müller, T. Davitashvili, M. Capek:** Data Curation, Writing - Review & Editing

**T. Yu, A. Vardazaryan, C.I. Nwoye, N. Padoy, X. Liu, E.J. Lee, C. Disch, H. Meine, T. Xia, F. Jia, S. Kondo, W. Reiter, Y. Jin, Y. Long, M. Jiang, Q. Dou, P. A. Heng, I. Twick, K. Kirtac, E. Hosgor, J.L. Bolmgren, M. Stenzel, B. von Siemens:** Data Analysis, Software, Writing - Review & Editing

**H.G. Kenngott, L. Maier-Hein, S. Speidel, B.P. Müller-Stich:** Conceptualization, Challenge organisation, Writing - Review & Editing, Supervision, Resources, Funding acquisition

## Acknowledgements

We thank T. Nguyen (team IGITech), L. Zhao, Z. Ge, H. Sun, D. Xie (team HIKVision) and M. Guo, C. Gong, D. Liu (team VIE-PKUE) for participation in the challenge. They were not listed as coauthors, because they did not respond to the requested authors confirmation, but agreed to publish their results when participating in the challenge.

# Funding

This work was supported by the National Center for Tumor Diseases (NCT) Heidelberg within the Cancer-Therapy-Program „Surgical Oncology", by the German Federal Ministry of Economic Affairs and Energy within project "InnOPlan" (grant number BMWI 01MD15002E), by the German Federal Ministry of Health within project "Surgomics" (grant number BMG 2520DAT82D), by the German Research Foundation DFG within the Cluster of Excellence EXC 2050: "Centre for Tactile Internet with Human-in-the-Loop (CeTI)" (project number 390696704), by the German Academic Exchange Service (DAAD) with a scholarship for T. Davitashvili for studying medicine in Germany (Scholarship program ID: deutsche Auslandsschulen, 2017 (57314589)), by the Helmholtz Imaging Platform (HIP), a platform of the Helmholtz Incubator on Information and Data Science, by Agence Nationale pour la Recherche (ANR-16-CE33-0009, ANR-10-IAHU-02) and Banque Publique d'Investissement (BPI CONDOR), by the Richard und Annemarie Wolf-Stiftung, by the Guangdong Key Area Research and Development Program (2020B010165004), by the Shenzhen Key Basic Science Program (JCYJ20180507182437217), by the Hong Kong RGC TRS Project No.T42-409/18-R.

None of the funding sources had influence on study design, the collection, analysis and interpretation of data, the writing of the report or the decision to submit the article for publication.

# Appendices

## Appendix A - Annotation rules

In the following, additional details on the annotation protocol for phase, action, instrument, and skill annotation as described in the "materials & methods" section of the main article is presented. The annotations are at least two frames long, as the video annotation research tool Anvil does not allow the independent annotation of a single frame.

### Phase annotation rules

The phase annotation includes seven phases: preparation (P0), calot triangle dissection (P1), clipping and cutting (P2), gallbladder dissection (P3), gallbladder packaging (P4), cleaning and coagulation (P5) and gallbladder retraction (P6). The phases do not necessarily occur in a fixed order.

The preparation phase (P0) begins as soon as the camera is inserted into the abdomen for the first time and the optical trocar is no longer visible in the image. This phase includes orientation in the patient's abdomen and placement of additional trocars for instrument insertion.

Calot triangle dissection (P1) begins as soon as an instrument appears in the image. The phase includes the dissection of connective tissue and fat around the gallbladder and the nearby abdominal cavity to reach the calot triangle and the preparation of the cystic artery and the cystic duct. One main cystic artery and one cystic duct were assumed for each video to reduce complexity, so that clipping and cutting (P2) was annotated for only one specific artery and one cystic duct, but not for additional vessels.

The phase clipping and cutting (P2) begins with the appearance of a clipper, which clips the cystic artery or the cystic duct before cutting. The phase does not begin when a clipper appears to stop bleeding or to clip other vessels. This phase may switch with the previous phase (P1). The change is defined by the appearance of a preparation instrument, for example the electric hook or the overholt, which actually dissects. The change does not take place if an instrument that is not dissecting appears in the image. In case that the electric hook or scissors start dissecting immediately after cutting the vessels, the phase P1 begins as soon as tissue is cut or cauterized.

Gallbladder dissection (P3) begins after the clipping and cutting (P2) as soon as the preparation scissors or the electric hook touches the gallbladder. The phase also begins with accidental contact, without cutting intention. It includes the clipping of accidentally injured vessels and accessory biliary arteries which are not defined as the main cystic vessels.

The gallbladder packaging phase (P4) begins as soon as the specimen bag enters the picture. This phase includes the recovery of the gallbladder as well as any gallstones. In the case of collecting spilled gallstones after the gallbladder has been packed in the specimen bag, the phase is maintained as long as the stones are placed in the bag and it is not closed. This phase ends with the complete closure of the bag or if the focus is diverted from the specimen bag to start the following phase.

Cleaning and coagulation (P5) begins as soon as the focus of the camera is averted from the specimen bag to check for tissue damage, coagulation begins using electricity, or drainage enters the picture. If the surgical site is checked and no cleaning or coagulation takes place, the phase does not begin. If the retraction of the gallbladder takes place before cleaning and coagulation, the phase begins as soon as the liver falls into the camera focus.

The gallbladder retraction phase (P6) begins when the specimen bag is last grasped to remove it from the abdomen.

The end of the operation is defined by the optical trocar taking 50 or more percent of the image as the camera is pulled out of the abdomen for the last time.

## Action annotation rules

The action annotation includes the four actions grasp (A0), hold (A1), cut (A2) and clip (A3). In connection with this, the performer of the action was annotated as the left hand of the surgeon, right hand of the surgeon or hand of the assistant.

The annotation of the action grasp (A0) started as soon as the grasper began to close the instrument and ended with its complete closure. It did not matter whether the grasper grasped tissue or closed without grasping tissue, for example, when the surgeon missed the tissue. Thus, the angle of closure of the instrument at the end of the action varied depending on the thickness of the grasped tissue between the two tips of the grasper. This action is only performed by instruments with the name grasper (I0-3, I10, I16, I18, I19) or forceps (I14, I15, I17). This action is not performed by the overholt (I4), since its primary function in laparoscopic cholecystectomy is to separate tissue bluntly by spreading.

The annotation of the action hold (A1) started after the action grasp, if the grasper successfully grasped tissue. It ended as soon as the grasper began to release its grip or when tissue started to slip out of the grip. In this context the end of the action hold is not always apparent, for example when the instrument releases gripped tissue outside of the camera view, is concealed by smoke or is pulled out of the abdomen while holding, as is often the case with the retraction of the specimen bag in the final phase. For this reason, the started action is annotated, even if the performing instrument is no longer visible. To recognize the end of the action correctly, the tension of the tissue must be considered. For example, if the assistant lifts the gallbladder until the performed instrument can no longer be seen, it is necessary to look for a decrease in tension due to the gallbladder tissue slipping out of the grip and to recognise the associated need for regrasp. In case of pulling the instrument out of the abdominal cavity while holding something, for example the specimen bag, gallstones or misplaced clips, the action hold is annotated until the held object disappears from the camera view and is pulled out of the abdomen.

The annotation of the action cut (A2) started as soon as the cutting instrument began to close around tissue and ended with the complete closure of the cutting instrument. In contrast to action grasp (A0), closing a cutting instrument without tissue was not annotated as cutting. This action also includes the tearing of the tissue by the scissors.

The annotation of the action clip (A3) started as soon as the clipper beganto close around tissue and ended with the clipper beginning to release its grip after the application of a clip.

## Instrument annotation rules

21 performing instruments were annotated and additionally divided into the seven categories grasper (IC0), clipper (IC1), coagulation instruments (IC2), scissors (IC3), suction-irrigation (IC4), specimen bag (IC5) and stapler (IC6).

For instrument presence, an instrument was considered visible as soon as its characteristic instrument tip appeared in the image. The annotation continues when the tip disappears later and only the shaft of the instrument remains visible. An example is the disappearance of the instrument tip of the electric hook behind tissue during dissection. Most importantly, the shaft should be clearly associated with an instrument tip.

If the instrument shaft enters the picture without its tip having been visible before, it is referred to as the undefined instrument shaft (I30), because even a human annotator would have difficulties recognizing a particular instrument due to the identical looking shaft. Three exceptions are the suction-irrigation (I12), stapler (I20) and the clippers (I8, I9), as these instruments have characteristic shafts.

One special case for the annotation of instrument presence is the concealment of an instrument by smoke resulting from coagulation. If the instrument tip and shaft are completely or partially obscured for a short time, the instrument is annotated as long as it was not removed from the camera view in the meantime.

## Skill annotation rules

As outlined in the methods section, the skill assessment was conducted using a modified GOALS score. It has been validated for video assessment of laparoscopic skills, including the five domains depth perception (S1), bimanual dexterity (S2), efficiency (S3), tissue handling (S4) and autonomy (Vassiliou et al., 2005). The item "autonomy" was omitted in our study, because a valid assessment based solely on intraabdominal video alone is not possible without information about what was spoken during the operation or how much assistance was provided by a senior surgeon. The difficulty of the operation (S5) was additionally annotated based on Chang's adaptation of the GOALS-score (Chang et al., 2007). Here, parameters such as inflammatory signs, adhesions and individual anatomical conditions were used to objectify the assessment of the skill. Thus, the skill assessment in this study included five ranking components. Skill was annotated for the complete operation and additionally for phases calot triangle dissection (P1) and gallbladder dissection (P3).

# Appendix B - Detailed challenge results

| | metrics | P0 | P1 | P2 | P3 | P4 | P5 | P6 | average |
|---|---|---|---|---|---|---|---|---|---|
| CAMI-SIAT | recall | 78,11% | 91,29% | 0,00% | 69,53% | 0,00% | 70,94% | 0,00% | 44,27% |
| | precision | 86,75% | 73,80% | 0,00% | 56,28% | 0,00% | 46,90% | 0,00% | 37,68% |
| | F1 | 78,19% | 79,57% | 0,00% | 60,01% | 0,00% | 52,75% | 0,00% | 38,65% |
| CAMMA 1 | recall | 99,88% | 81,28% | 61,63% | 83,01% | 65,47% | 43,05% | 75,97% | 72,90% |
| | precision | 87,80% | 84,11% | 69,82% | 80,58% | 49,87% | 64,19% | 62,73% | 71,30% |
| | F1 | 92,83% | 80,10% | 55,92% | 80,24% | 54,00% | 45,33% | 65,43% | 68,78% |
| CAMMA 2 | recall | 98,15% | 86,76% | 62,70% | 74,78% | 70,70% | 21,54% | 69,54% | 69,17% |
| | precision | 95,36% | 84,83% | 59,02% | 75,20% | 30,84% | 50,92% | 52,99% | 64,17% |
| | F1 | 96,59% | 84,97% | 60,18% | 73,79% | 38,80% | 25,50% | 58,32% | 63,60% |
| CUHK | recall | 100,00% | 85,98% | 41,56% | 74,22% | 47,94% | 50,71% | 68,52% | 66,99% |
| | precision | 86,17% | 83,45% | 60,17% | 76,87% | 66,27% | 49,86% | 52,33% | 67,88% |
| | F1 | 91,99% | 83,74% | 47,89% | 74,66% | 51,37% | 49,41% | 55,83% | 64,98% |
| HIKVision | recall | 88,49% | 86,50% | 46,92% | 74,88% | 45,90% | 58,83% | 62,03% | 66,22% |
| | precision | 76,15% | 81,26% | 78,26% | 83,80% | 58,54% | 56,42% | 55,44% | 69,98% |
| | F1 | 79,39% | 82,87% | 57,06% | 77,40% | 50,19% | 54,75% | 56,00% | 65,38% |
| IGIMedicalTechnologies | recall | 10,68% | 7,75% | 1,32% | 78,11% | 52,50% | 36,81% | 34,34% | 31,65% |
| | precision | 57,57% | 56,63% | 56,32% | 27,13% | 41,17% | 25,64% | 39,04% | 43,36% |
| | F1 | 15,86% | 13,29% | 2,53% | 38,22% | 41,67% | 25,43% | 30,53% | 23,93% |
| MEVIS | recall | 82,15% | 79,36% | 46,25% | 60,62% | 50,48% | 54,48% | 65,77% | 62,73% |
| | precision | 67,76% | 84,82% | 72,19% | 77,40% | 43,84% | 51,88% | 33,28% | 61,60% |
| | F1 | 70,31% | 79,39% | 49,33% | 66,80% | 42,48% | 50,71% | 42,05% | 57,30% |
| NCT | recall | 61,26% | 90,84% | 50,89% | 82,95% | 26,36% | 16,10% | 9,80% | 48,31% |
| | precision | 97,12% | 78,58% | 77,32% | 48,74% | 48,35% | 38,81% | 44,14% | 61,87% |
| | F1 | 69,97% | 83,11% | 58,88% | 59,41% | 32,53% | 18,86% | 14,75% | 49,00% |
| VIE-PKU | recall | 100,00% | 68,99% | 11,40% | 34,12% | 37,62% | 23,59% | 0,00% | 39,39% |
| | precision | 74,61% | 62,56% | 29,28% | 29,39% | 17,43% | 18,91% | 0,00% | 33,17% |
| | F1 | 84,47% | 57,43% | 16,09% | 30,11% | 22,25% | 18,96% | 0,00% | 33,29% |
| Wintegral | recall | 57,13% | 60,99% | 44,65% | 40,68% | 30,68% | 47,52% | 29,43% | 44,44% |
| | precision | 62,34% | 61,79% | 44,03% | 55,46% | 47,25% | 36,86% | 23,73% | 47,35% |
| | F1 | 56,99% | 57,83% | 41,51% | 46,29% | 33,00% | 38,05% | 23,61% | 42,47% |

**Phase recognition results.** Phase recognition results of all participants. Phases displayed are: preparation (P0), calot triangle dissection (P1), clipping and cutting (P2), gallbladder dissection (P3), gallbladder packaging (P4), cleaning and coagulation (P5) and gallbladder retraction (P6).

|  | metrics | A0 | A1 | A2 | A3 | average |
|---|---|---|---|---|---|---|
| CAMMA 1 | recall | 4,36% | 99,78% | 14,19% | 1,01% | 29,83% |
|  | precision | 2,60% | 73,61% | 0,43% | 0,11% | 19,19% |
|  | F1 | 2,91% | 84,50% | 0,83% | 0,18% | 22,10% |
| KonicaMinolta | recall | 0,00% | 93,16% | 0,00% | 0,15% | 23,33% |
|  | precision | 0,00% | 89,17% | 0,00% | 12,50% | 25,42% |
|  | F1 | 0,00% | 91,02% | 0,00% | 0,30% | 22,83% |
| NCT | recall | 5,31% | 99,99% | 0,00% | 0,70% | 26,50% |
|  | precision | 1,70% | 77,52% | 0,00% | 2,76% | 20,49% |
|  | F1 | 2,36% | 87,01% | 0,00% | 1,12% | 22,62% |
| VIE-PKU | recall | 0,00% | 85,00% | 0,00% | 0,47% | 21,37% |
|  | precision | 0,00% | 87,90% | 0,00% | 5,35% | 23,31% |
|  | F1 | 0,00% | 86,14% | 0,00% | 0,87% | 21,75% |
| Wintegral | recall | 0,00% | 90,99% | 0,00% | 2,12% | 23,28% |
|  | precision | 0,00% | 88,79% | 0,00% | 8,67% | 24,36% |
|  | F1 | 0,00% | 89,71% | 0,00% | 3,41% | 23,28% |

**Action recognition results.** Action recognition results of all participants. Actions displayed are: grasp (A0), hold (A1), cut (A2) and clip (A3).

|  | metrics | IC0 | IC1 | IC2 | IC3 | IC4 | IC5 | average |
|---|---|---|---|---|---|---|---|---|
| CareSyntax | recall | 77,81% | 69,75% | 80,26% | 53,80% | 32,65% | 66,68% | 63,49% |
|  | precision | 84,37% | 34,67% | 78,41% | 20,74% | 19,74% | 42,35% | 46,71% |
|  | F1 | 79,38% | 43,66% | 78,44% | 26,69% | 22,84% | 49,75% | 50,13% |
| CUHK | recall | 83,06% | 57,67% | 82,09% | 39,02% | 31,35% | 56,69% | 58,31% |
|  | precision | 87,73% | 87,49% | 88,97% | 79,89% | 30,27% | 79,94% | 75,71% |
|  | F1 | 84,84% | 66,97% | 85,25% | 47,74% | 30,03% | 62,87% | 62,95% |
| HIKVision | recall | 87,21% | 55,66% | 84,32% | 30,74% | 28,54% | 44,18% | 55,11% |
|  | precision | 82,96% | 75,28% | 85,53% | 68,35% | 28,21% | 70,83% | 68,53% |
|  | F1 | 84,79% | 63,55% | 84,37% | 38,40% | 26,75% | 51,34% | 58,20% |
| IGIMedicalTechnologies | recall | 88,54% | 5,82% | 85,41% | 0,40% | 20,30% | 41,07% | 40,25% |
|  | precision | 72,17% | 74,27% | 66,56% | 28,06% | 25,84% | 64,14% | 55,17% |
|  | F1 | 79,20% | 10,11% | 73,80% | 0,77% | 21,15% | 48,12% | 38,86% |
| KonicaMinolta | recall | 84,20% | 66,15% | 83,12% | 52,49% | 31,38% | 59,20% | 62,76% |
|  | precision | 87,61% | 80,42% | 88,25% | 67,35% | 28,08% | 67,91% | 69,94% |
|  | F1 | 85,71% | 71,33% | 85,34% | 51,65% | 28,16% | 60,69% | 63,82% |
| NCT | recall | 100,00% | 100,00% | 91,78% | 0,00% | 5,36% | 99,11% | 66,04% |
|  | precision | 57,40% | 4,05% | 48,55% | 0,00% | 6,82% | 9,01% | 20,97% |
|  | F1 | 72,62% | 7,67% | 62,76% | 0,00% | 5,37% | 16,25% | 27,45% |
| VIE-PKU | recall | 94,16% | 26,08% | 89,53% | 24,64% | 27,51% | 82,79% | 57,45% |
|  | precision | 66,41% | 28,25% | 63,85% | 17,49% | 12,46% | 18,34% | 34,47% |
|  | F1 | 77,53% | 21,70% | 73,37% | 14,56% | 14,42% | 29,23% | 38,47% |
| Wintegral | recall | 70,47% | 23,56% | 67,37% | 13,63% | 13,67% | 37,27% | 37,66% |
|  | precision | 79,16% | 55,78% | 82,21% | 18,45% | 24,79% | 59,08% | 53,25% |
|  | F1 | 73,41% | 30,63% | 70,85% | 13,91% | 16,71% | 44,05% | 41,59% |

**Instrument presence detection results.** Instrument presence detection results of all

> participants. Instrument categories displayed are: grasper (IC0), clipper (IC1), coagulation instruments (IC2), scissors (IC3), suction-irrigation (IC4) and specimen bag (IC5). The stapler category (IC6) is not displayed as it was not present during the test data set.

# Vitae

**Martin Wagner** is a general surgeon at Heidelberg University Hospital. He is head of the research group on artificial intelligence and cognitive robotics within the division of minimally invasive and robot-assisted surgery in the department of surgery. His group brings together surgeons, medical students, computer scientists, roboticists, and designers to create novel solutions for clinical problems in surgery. Together with cooperation partners from academia and industry they seek to leverage the potential of machine learning in helping surgeons to choose the best treatment option for the individual patient and perform the surgical treatment in the best possible way.

**Patrick Heger** is a general surgeon at Heidelberg University Hospital. He is a scientist and trial physician at the clinical study center (KSC) and takes part in the conduction of clinical trials at the Department of Surgery. Furthermore, he is the principal investigator of several trials there. Additionally, he is a member of the Study Center of the German Society of Surgery (SDGC).

**Lars Mündermann** is an employee at KARL STORZ. He is head of the data assisted solutions group within corporate research & technology. His group works on the evaluation and assessment of new technologies for computer-assisted surgery.

**David M.Lubotsky** studied at the LMU in Munich and received his BSc degree in physics. Currently he is a medical student at Heidelberg University and research assistant within the department of general, visceral and transplantation surgery of Heidelberg University Hospital.

**Annika Reinke** studied applied mathematics at the University of Lübeck, Germany, with a focus on medical image analysis. In 2017 she joined the division of Computer Assisted Medical Interventions at the German Cancer Research Center (DKFZ) to work on scientific benchmarking and validation of AI algorithms, since 2020 as her PhD topic. Having published disruptive findings on biomedical image analysis challenges in Nature Communications, she is a founding member of the initiative of Biomedical Image Analysis ChallengeS (BIAS). She further serves as an active member in several working groups with focus on open science, which led to several high-ranked publications.

**Xinyang Liu** is a Staff Scientist at Sheikh Zayed Institute for Pediatric Surgical Innovation, Children's National Hospital of Washington DC. He received his PhD degree from Florida State University in 2010. He worked as a Postdoctoral Fellow at Johns Hopkins Hospital and Brigham and Women's Hospital. His research interests include medical augmented reality, computer assisted surgery and medical image analysis.

**Hans Meine** received a PhD from the University of Hamburg in 2008 for his fundamental research on image segmentation methods at the department for computer science. Since 2011, he is applying image analysis to various medical image modalities at Fraunhofer MEVIS, where he is nowadays coordinating image analysis and deep learning developments across projects. From 2015-2021, he had a second position at the University of Bremen for research and teaching.

**Satoshi Kondo** is an associate professor at Muroran Institute of Technology, Japan. He received his B.S., M.S. and Ph.D degrees at Osaka Prefecture University in 1990, 1992 and 2005, respectively. He was with Panasonic Corporation and KonicaMinolta Inc. His research interests are in the fields of computer vision.

**Franziska Mathis-Ullrich** is Assistant Professor for Medical Robotics at the Karlsruhe Institute of Technology (KIT), Germany. Her research focus is on cognition controlled robotics for minimally-invasive surgery. She received her B.Sc. (2009), M.Sc. (2012) and Ph.D. (2017) from ETH Zurich. Since 2019, she is head of the Health Robotics and Automation Laboratory at KIT. Franziska Mathis-Ullrich has received multiple academic awards (IEEE ICRA Best Paper Award, 2014; IEEE BioRob Best Student Paper Award, 2016; winner of ICRA Microassembly Challenge, 2014 & 2015). In 2017 she made it onto the prestigious Forbes "30 under 30" list.

**Lena Maier-Hein** is a full professor at Heidelberg University (Germany) and affiliated professor to LKSK institute of St. Michael's Hospital (Toronto, Canada). At the German Cancer Research Center (DKFZ) she is managing director of the "Data Science and Digital Oncology" cross-topic program and head of the division Computer Assisted Medical Interventions (CAMI). Her research concentrates on machine learning-based biomedical image analysis with a specific focus on surgical data science, computational biophotonics and validation of machine learning algorithms.

**Stefanie Speidel** is a full professor for "Translational Surgical Oncology" at the National Center for Tumor Diseases (NCT) Dresden and one of the speakers of the DFG Cluster of Excellence "Centre for Tactile Internet with Human-in-the-Loop (CeTI)". Her current research interests focus on machine learning for computer- and robot-assisted surgery in the context of the future digital operating room.

**Sebastian Bodenstedt** is a postdoctoral researcher at the National Center for Tumor Diseases (NCT) Dresden and member of the Centre for Tactile Internet with Human-in-the-Loop (CeTI)". His current research interests focus on developing machine-learning and computer-vision methods for the surgical environment.